\documentclass{newclass}

\usepackage{hyperref}
\usepackage{amsmath}
\usepackage{amssymb}
\usepackage{graphicx}

\usepackage{subfigure}     

\begin{document}

\chapter{Experimental Methods for Generating Two-Dimensional Quantum Turbulence in Bose-Einstein Condensates}

\author[K. E. Wilson, E. C. Samson, Z. L. Newman, T. W. Neely, and B. P. Anderson]{K. E. Wilson, E. C. Samson\footnote{Current address: School of Physics, Georgia Institute of Technology, Atlanta, Georgia 30332, USA.}, Z. L. Newman, T. W. Neely\footnote{Current address: School of Mathematics and Physics, University of Queensland, Qld 4072, Australia.}, and B. P. Anderson}

\address{College of Optical Sciences, University of Arizona,\\
1630 E. University Blvd., Tucson, Arizona 85711, USA}

\begin{abstract}

Bose-Einstein condensates of dilute gases are well-suited for investigations of vortex dynamics and turbulence in quantum fluids, yet there has been little experimental research into the approaches that may be most promising for generating states of two-dimensional turbulence in these systems.  Here we give an overview of techniques for generating the large and disordered vortex distributions associated with two-dimensional quantum turbulence.  We focus on describing methods explored in our Bose-Einstein condensation laboratory, and discuss the suitability of these methods for studying various aspects of two-dimensional quantum turbulence.  We also summarize some of the open questions regarding our own understanding of these mechanisms of two-dimensional quantum turbulence generation in condensates. We find that while these disordered distributions of vortices can be generated by a variety of techniques, further investigation is needed to identify methods for obtaining quasi-steady-state quantum turbulence in condensates.
\end{abstract}

\body


\section{Introduction}\label{KW_sec1}

Dilute-gas Bose-Einstein condensates (BECs) present unique opportunities for studying quantum fluid dynamics  due to the wide range of experimental tools available for probing and manipulating condensates, and the quantitative accuracy available with theoretical and numerical approaches \cite{Pet2008}.   Experimentally, magnetic and laser fields can be arranged to create a wide variety of trapping configurations and geometries. Superfluidity hallmarks such as persistent currents and quantized vortices \cite{Don1991.book.Vortices,Anderson2010} may be precisely generated and their dynamics probed, and numerous measurement techniques allow for observations of density and quantum phase distributions.  Regarding simulations of superfluid dynamics in BECs, common numerical approaches utilize the Gross-Pitaevskii equation (GPE), a nonlinear Schr\"{o}dinger equation that incorporates a mean-field approximation to represent the BEC with an order parameter, the familiar macroscopic condensate wavefunction \cite{Pit2003.book.Bose}.  Numerous variants of the GPE such as those that model finite-temperature environments extend the predictive accuracy of simulations \cite{Blakie08a}.  Taken together, these features allow for highly synergetic experimental, numerical, and theoretical explorations of superfluid dynamics.

In this respect, Bose-Einstein condensates may be ideally suited for investigating quantum turbulence and the dynamics of interacting quantized vortices in compressible quantum fluids; for recent reviews of this subject see Refs.~\cite{Tsubota09a} and \cite{Tsubota2012} and the references therein.  In superfluids, vorticity appears in the form of superfluid-free vortex cores about which there is quantized flow circulation \cite{Don1991.book.Vortices}.  Whether in superfluid helium or trapped BECs, quantum turbulence is often phenomenologically described as a tangled distribution of these quantized vortices throughout the fluid \cite{Feynman1955,QVDSF}, and recent work with superfluid helium provides fascinating visual evidence for these vortex distributions \cite{Paoletti08a, Paoletti2011}. 

In BECs, quantum turbulence is relevant to early investigations on the formation of large vortex lattices \cite{Abo2001.Sci292.5516} and routes to vortex nucleation by dynamical instabilities \cite{Mad2001.PRL86.4443,Sin2001.PRL87.190402}, and has been experimentally observed and explicitly addressed in these contexts \cite{Mad2000.PRL84.806,Che2000.PRL85.2223,Ram2001.PRL87.210402,Sch2004.PRL93.210403}.  More recent experiments have been principally dedicated to the study of quantum turbulence phenomena and the related vortex distributions \cite{Henn09a,Sem2011.LPL8.691,Car2012.JLTP2012.49,Neely2012}.  Images of vortex tangles in BECs have also been experimentally obtained \cite{Ram2001.PRL87.210402,Henn09a}.  
Yet while BECs readily permit vortex observations, other aspects of turbulence remain significant experimental challenges within the BEC field. Notably, direct measurement of kinetic energy spectra is an open problem, whereas such measurements of liquid helium have allowed direct comparison of classical and quantum fluid turbulent energy spectra \cite{Maurer1998a}.  The different approaches and systems for studying quantum turbulence are therefore complementary, with BEC research playing a potentially important role in understanding the broad subject of turbulence, particularly in the context of compressible quantum fluids.
  
More specifically, BECs are now beginning to contribute to the development of an understanding of \emph{two-dimensional} (2D) turbulence \cite{Kra1967.PF10.1417,Lei1968.PF11.671,Bat1969.PF12.II233} in compressible quantum fluids.  In the basic manifestation of 2D classical turbulence (2DCT), energy and enstrophy\footnote{
Over a 2D surface $S$, the enstrophy $\Omega = \int_S dS\, |\omega|^2$ is a measure of
  the vorticity $\omega = [ \nabla \times \vec{v} ]_z$ about the $z$ axis, where $\vec{v}$
  is the fluid velocity in the plane normal to $z$.} 
 are injected into a fluid at a small length scale, and patches of vorticity merge and lead to the development of large-length-scale flow.  This is opposite to the dynamics of the familiar energy cascade of three-dimensional (3D) turbulence in a classical fluid, in which large-scale flows continuously decay to smaller-scale flows that are eventually dissipated by viscous damping.  This \emph{inverse} energy cascade of 2DCT has been studied for decades and numerous excellent reviews exist on 2DCT \cite{Sommeria01a,Tabeling2002,Boffetta12a}.

\subsection {Experimental study of two-dimensional quantum turbulence in BECs}

In order to study two-dimensional \emph{quantum} turbulence (2DQT) in a BEC, one might envision subjecting the BEC to continuous injection of energy and vorticity.  Dissipation of energy may also exist, perhaps by thermal damping that could induce vortices to exit the BEC at its boundary, or perhaps due to vortex-antivortex annihilation.  A balance between forcing and dissipation can be envisioned to lead to a statistically steady degree of turbulence in a BEC.  We might picture such a state as an approximately constant mean number of vortices in the BEC, since enstrophy in a quantum fluid is proportional to the number of vortices \cite{Numasato10b,Bradley2012}.  With the likelihood of future experimental methods for observing vortex dynamics and measuring energy spectra, we further imagine a scenario in which 2DQT can be experimentally and numerically studied, linked with new analytical approaches, and compared with 2DCT phenomena.  The prospects for the development of a new understanding of at least this one aspect of the exceptionally challenging topic of turbulence are tantalizing motivations for pursuit of 2DQT research.

Unfortunately there are a number of experimental challenges that must first be overcome.  Significantly, an isolated, trapped BEC subjected to continuous injection of energy would eventually become depleted of atoms as the system rethermalized.   It appears that a BEC experimentalist's primary hope in this respect is to develop methods for studying 2DQT that are quasi-steady, where steady forcing and dissipation rates are balanced over finite timescales during which the BEC is not significantly depleted of atoms.  Experimentally, the path to studying decaying 2DQT is perhaps much simpler, since this topic is more concerned with the dynamics of a BEC after forcing has stopped. 

While it should not be surprising that disordered vortex distributions in a highly oblate BEC are relatively simple to obtain by a variety of experimental techniques, as we discuss in this article, how best to obtain a quasi-steady state of forced 2DQT remains an open problem.  We have begun to empirically tackle only the forcing aspect of this problem, although simulations are also beginning to address this issue \cite{Neely2012,Bradley2012,Fujimoto2011,White2012a}.  Several of our methods for driving large numbers of vortices into highly oblate BECs are described below.  We generally associate these states with 2DQT in the sense that the disordered vortex distributions may be considered a 2D phenomenological equivalent of the vortex tangle that exists in 3D quantum turbulence.  We do not provide any evidence that such disordered vortex distributions conform to particular kinetic energy spectra or display vortex dynamics that may be expected in fully developed 2DQT; discussions of links between vortex distributions, vortex dynamics and aggregates, and kinetic energy spectra in BEC 2DQT can be found elsewhere \cite{Neely2012,Bradley2012,White2012a,Now2012.PRA85.043627,Schole2012a}.   We also note that in the highly oblate trapping limit, BEC confinement in one spatial direction is much stronger than the trapping in the two orthogonal directions.  This limit is not strictly necessary for studying 2DQT, but vortex excitations in the form of Kelvin waves are suppressed and vortices can be accurately approximated as having point-like dynamics in a plane within this limit \cite{Rooney11a}.

In the following, we first describe some of the key phenomenological features of 2DCT, and summarize only a few of the theoretical advances in 2DQT, some of which have been aimed at uncovering the similarities and differences between the classical and quantum systems.  This article is not intended as a review of the state of theoretical understanding of 2DQT in BECs; for discussions of theoretical aspects of 2DQT we direct the reader to Refs.~\cite{Neely2012,Numasato10b,Bradley2012,White2012a,Now2012.PRA85.043627,Schole2012a,Nowak2011a} and the references contained therein.   The study of 2DQT, particularly in compressible quantum fluids such as BECs, is still young and a full range of characteristics remains to be explored.  Our aim is to give an overview of the types of observations that may be desirable in 2DQT BEC experiments, then to describe our experimental approach to creating and probing 2DQT in highly oblate BECs.   We present a range of tools available for exciting the BEC using laser beams and magnetic fields.  The bulk of the remainder of this article is devoted to describing various specific methods for driving vortices into a highly oblate BEC, and we show example images of BECs in such excited states.  We conclude by briefly summarizing some of the open experimental challenges in generating and probing 2DQT in BECs.


\section{Overview of Two-Dimensional Turbulence}\label{KW_sec2}


From a phenomenological standpoint, Lesieur describes classical turbulence of continuum flows as having non-deterministic flow details, rapid mixing relative to molecular diffusion, and flow characteristics distributed among and interacting over a wide range of length scales \cite{Les2008.Turbulence}; see also Sommeria's similar characterization of turbulence \cite{Sommeria01a}. Such a description encompasses 2D turbulence, where flow dynamics vary primarily over two spatial coordinates ($x$ and $y$); flow in a third direction ($z$) may exist uniformly throughout the $x$-$y$ plane or there may be negligible flow along $z$, hence a 3D fluid system may display 2D turbulence.  Building from this general description, the further characteristics of 2D turbulence are markedly different from the characteristics of 3D turbulence.  As noted earlier, one such difference is the direction of energy flux between length scales: in 2D flows, small-scale forcing can lead to the development of large-scale flows.

\subsection{Key ideas of 2D turbulence}\label{KW_sec2_1}

Given their disordered and unpredictable nature, and challenges in discerning details of flow dynamics, turbulent flows are often characterized by the statistical properties of the system such as the kinetic energy density $E(k)$ over wavenumber $k$, which describes the distribution of kinetic energy among the length scales of the system. In 1967, Kraichnan \cite{Kra1967.PF10.1417} found the existence of two inertial ranges associated with distinct kinetic energy spectra power laws, joined by the wavenumber  associated with energy and enstrophy forcing.   The first range extends from the forcing length scale to larger length scales (or smaller wavenumbers), and  displays a kinetic energy spectrum $E(k) \propto k^{-5/3}$. This spectral distribution has the same form discovered by Kolmogorov in 1941 \cite{Kolmogorov1941} in the context of 3D fluids, but in 2D this spectrum corresponds to an inverse energy cascade.   The second range extends from the forcing scale to smaller length scales (or larger wavenumbers) and the kinetic energy spectrum approximately corresponds to $E(k) \propto k^{-3}$.

This two-component spectrum is a classic feature of 2D turbulence, and is further described in reviews of 2D turbulence~\cite{Sommeria01a,Tabeling2002,Boffetta12a}.   Physically the inverse energy cascade corresponds to the aggregation of vorticity such that small-scale forcing leads to the growth of large-scale vorticity until flows are of the order of the size of the system.  At this scale dissipation of some form occurs, possibly due to frictional damping at the container's walls.  

Prior to the development of 2D turbulence theory in the 1960s \cite{Kra1967.PF10.1417,Lei1968.PF11.671,Bat1969.PF12.II233}, Onsager \cite{Ons1949.NC6s2.279,Eyi2006.RMP78.87} approached turbulent flow in 2D inviscid fluids by focusing on the dynamics of point-like centers of vorticity, or point vortices.  A collection of point vortices may be constructed to approximately represent classical continuum flows, or may quite accurately describe vortex distributions in superfluids.  In fact, in the same conference proceeding in which Onsager described his theory of vortex states in 2D turbulent flows, he also first postulated the existence of quantized vorticity in superfluids, and implied that superfluids could be ideal testing grounds for his theoretical arguments of 2D turbulence \cite{Ons1949.NC6s2.279}.  Onsager argued that a system with a large number of vortices would maximize entropy when vortices of the same sense of circulation form aggregates, giving more room in a finite-volume phase space for the remainder of the vortex cores to distribute themselves and move with fewer constraints.  Onsager's predictions, which were the first hints of the existence of an inverse energy cascade, have yet to be directly confirmed experimentally in a superfluid.  However, numerical simulations of 2DQT involving forcing in BECs are indeed finding evidence for the existence of vortex aggregates that resemble Onsager's predictions \cite{Neely2012,Bradley2012}.


\subsection{Vortices and 2D Quantum Turbulence in BECs}\label{KW_sec2_2}

Based on the above descriptions of 2D turbulence, direct experimental challenges for the study of 2D turbulence in quantum fluids include: (i) reaching a quasi-steady state of forced 2DQT; (ii) observation of the aggregation of like-circulation quantized vortices in a superfluid in a state of forced 2DQT; and (iii) direct measurement of the kinetic energy spectra associated with such turbulent states.  These challenges all depend on the existence of means to generate 2DQT, so we now turn to the subject of vortices in highly oblate BECs, and then focus on describing methods to nucleate large disordered distributions of vortices  in these compressible quantum fluids.

A dilute-gas BEC is often adequately described in terms of an order parameter, or macroscopic wavefunction, of the form \cite{Pet2008}
\begin{equation}
\psi(\vec{r},t) = \sqrt{n(\vec{r},t)}e^{i \phi(\vec{r},t)},
\end{equation}
where $n(\vec{r},t)$ is the atomic density distribution of the BEC and $\phi(\vec{r},t)$ is the quantum phase profile.  The dynamics of the wavefunction $\psi$ are
governed in the mean-field approximation by the Gross-Pitaevski equation (GPE)
\begin{equation} 
i\hbar \frac{\partial \psi}{\partial t}\, = \, -\frac{\hbar ^2}{2m}\nabla^2\psi\,+\,V(\vec{r},t)\psi\,+\,g|\psi|^2\psi
\end{equation}  
where $g = 4\pi\hbar^2a/m$ characterizes the interaction strength between atoms, $a$ is the s-wave scattering length of interatomic interactions, $m$ is the atomic mass, and $V(\vec{r},t)$ is the potential due to the trap and other external perturbations.

From a hydrodynamic perspective, a dilute-gas BEC can behave as a tiny droplet of superfluid, where the local velocity of superfluid flow is proportional to the local gradient of the phase:
\begin{equation}
\vec{v} = \frac{\hbar}{m}\nabla\phi.
\end{equation}
Consequently, the velocity field of a BEC is irrotational, $\nabla \times \vec{v} = 0$, except at singularities of the quantum phase $\phi$.  These singularities are the superfluid-free cores of the vortices, about which fluid circulation is quantized \cite{Pet2008}.  A singly quantized vortex is associated with a 2$\pi$-phase winding about the core, and vortex positions and circulations can therefore be experimentally determined using atom interferometry techniques \cite{Mat1999.PRL83.2498,Che2001.PRA64.031601,Ino2001.PRL87.080402}, or by sequential images that reveal the dynamics of the vortices \cite{And2000.PRL85.2857,Nee2010.PRL104.160401,Fre2010.Sci329.1182,Mid2011.PRA84.011605}.  Most single-shot images of BECs, however, only reveal the location of the vortices and not their circulation. 

\begin{figure}
\begin{center}
\includegraphics[width=.4\linewidth]{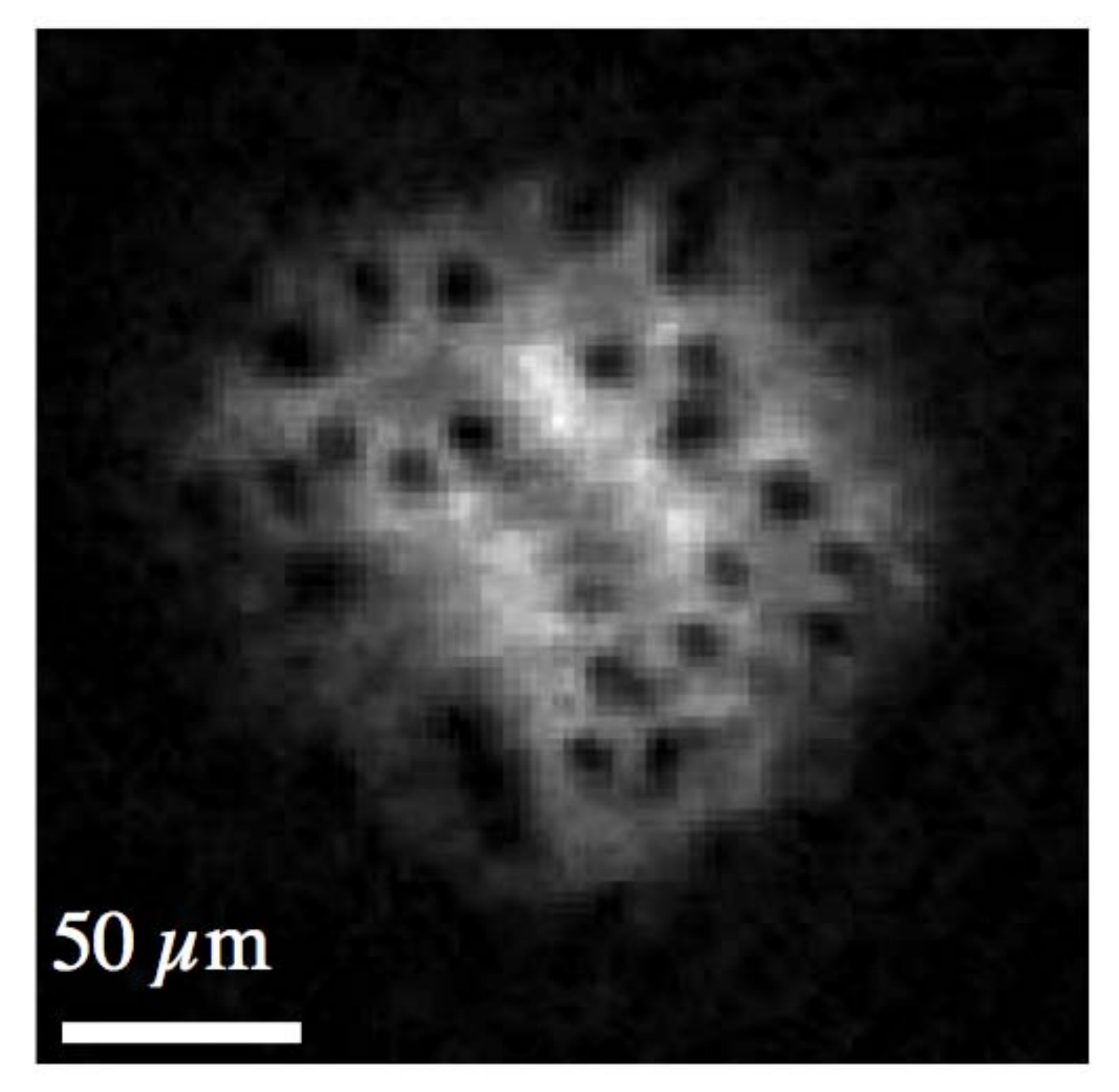}
\caption{A representative absorption image of an expanded oblate BEC with a resolvable disordered distribution of vortex cores.  The column density of the expanded cloud is proportional to grayscale, with lighter shades representing higher densities.  Vortices appear as the dark circular holes within the cloud.} 
\label{vortexfig}
\end{center}
\end{figure}

Vortex cores in a BEC have a diameter on the order of the healing length $\xi$ \cite{Pet2008}, which is typically sub-micron and much less than the wavelength of imaging light.   Release of the BEC from the trap followed by ballistic expansion of the atom cloud enables the vortices to expand to diameters large enough to be optically resolvable, and this method is the standard approach to observing vortices in a BEC. In a 3D trap, vortices may also bend with respect to the imaging axis \cite{Bre2003.PRL90.100403}, making them even more difficult to observe.  In two dimensions however, these Kelvin wave excitations are suppressed \cite{Rooney11a}.   As an example of the clarity that can be obtained in vortex detection, vortices in a highly oblate BEC are shown in Fig.~\ref{vortexfig}; the vortices are the circular dark regions within the BEC fluid, shown in grayscale with lightness of shade corresponding to integrated column density after ballistic expansion from the trap. 

In addition to isolated vortex cores, vortex dipoles may be found in highly oblate or 2D quantum fluids. These structures consist of two vortex cores of opposite circulation in close proximity.  Vortex aggregates, consisting of clusters of vortices of the same circulation, have also been experimentally observed in the form of multi-quantum vortex dipoles \cite{Nee2010.PRL104.160401}.   In the context of 2DQT in a BEC, however, to date only numerical simulations have observed vortex aggregates and their dynamics \cite{Neely2012}.  In BECs with a vector order parameter, other types of vortex structures may also be found; see Ref.~\cite{Anderson2010} for a summary of vortex experiments in BECs from their initial study in 1999 through 2010, including vortices in degenerate Fermi gases and multi-component BECs.

In the hydrodynamic regime, where quantum pressure can be neglected, and with vortices that are spaced far enough apart from each other that their core shapes are only very weakly distorted, the GPE formalism can be formally linked to the Navier-Stokes equation that plays a key role in classical fluid dynamics \cite{Les2008.Turbulence}.  By incorporating into the GPE damping due to atomic scattering with thermal atoms, a Reynold's number can also be defined \cite{Bradley2012}, although this has yet to be tested experimentally; see Ref.~\cite{Bradley2012} for a discussion of the GPE -- Navier-Stokes equation correspondence.  Given this correspondence, it should not be surprising that the concepts of turbulence may be investigated using BECs and the variants of the GPE.  

Much like the vortex tangle that characterizes quantum turbulence in three dimensions \cite{Feynman1955}, we would like to similarly physically characterize a state of 2DQT in a BEC. We consider the simple picture of a disordered distribution of numerous vortex cores.  Such a distribution would satisfy the phenomenological criteria of turbulence stated earlier.  First, a large disordered distribution of vortices would have non-deterministic and chaotic flow dynamics \cite{Aref1983}.  Second, numerical simulations have shown that such distributions can rapidly distribute vorticity \cite{Neely2012,Bradley2012} or particles \cite{Wan2007.JLTP149.65}, even if these are initially concentrated in a small region, thereby satisfying the rapid mixing criterion of turbulence.  Finally, such a distribution of vortices can display a large range of inter-vortex separations with no characteristic length scale; in contrast, a vortex lattice has a well-defined length scale of the spacing of the vortices in the lattice \cite{Bradley2012}.   We therefore interpret a disordered 2D distribution of numerous vortex cores as a phenomenological picture that corresponds to 2DQT.  Such a picture can be made more quantitative, for example by invoking the methodology and results of Onsager \cite{Ons1949.NC6s2.279} or vortex position statistics, as has been investigated numerically \cite{Bradley2012,White2012a,Novikov1975}.  Nevertheless, our general experimental goal is to create states that consist of many vortices distributed throughout highly oblate BECs with no discernible or well-defined arrangement.

As noted above, a primary factor distinguishing 2D and 3D turbulence is the suppression of significant flow dynamics in one spatial direction, relative to the flow dynamics in the plane normal to that direction.  In the case of quantum turbulence, this corresponds to the suppression of Kelvin waves \cite{Rooney11a}.  It may then be possible to study 2DQT in 3D BECs, and it should be noted that quasi-2D BECs \cite{Had2006.Nat441.1118} or even highly oblate BECs are not strictly needed for studies of 2DQT.   On the other hand, even in highly oblate BECs, vorticity can decay due to vortex-antivortex annihilation.  It is therefore possible that enstrophy need not be conserved and that a direct energy cascade may appear, as has been recently noted in numerical simulations of decaying 2DQT \cite{Numasato10b}.  In a practical sense however, we expect that 2DQT is more readily achievable in highly oblate geometries.    Our experimental focus is therefore on generating large disordered distributions of vortices in highly oblate BECs.


\section{Experimental Setup}\label{KW_sec3}

\begin{figure}
\begin{center}
\includegraphics[width=4in]{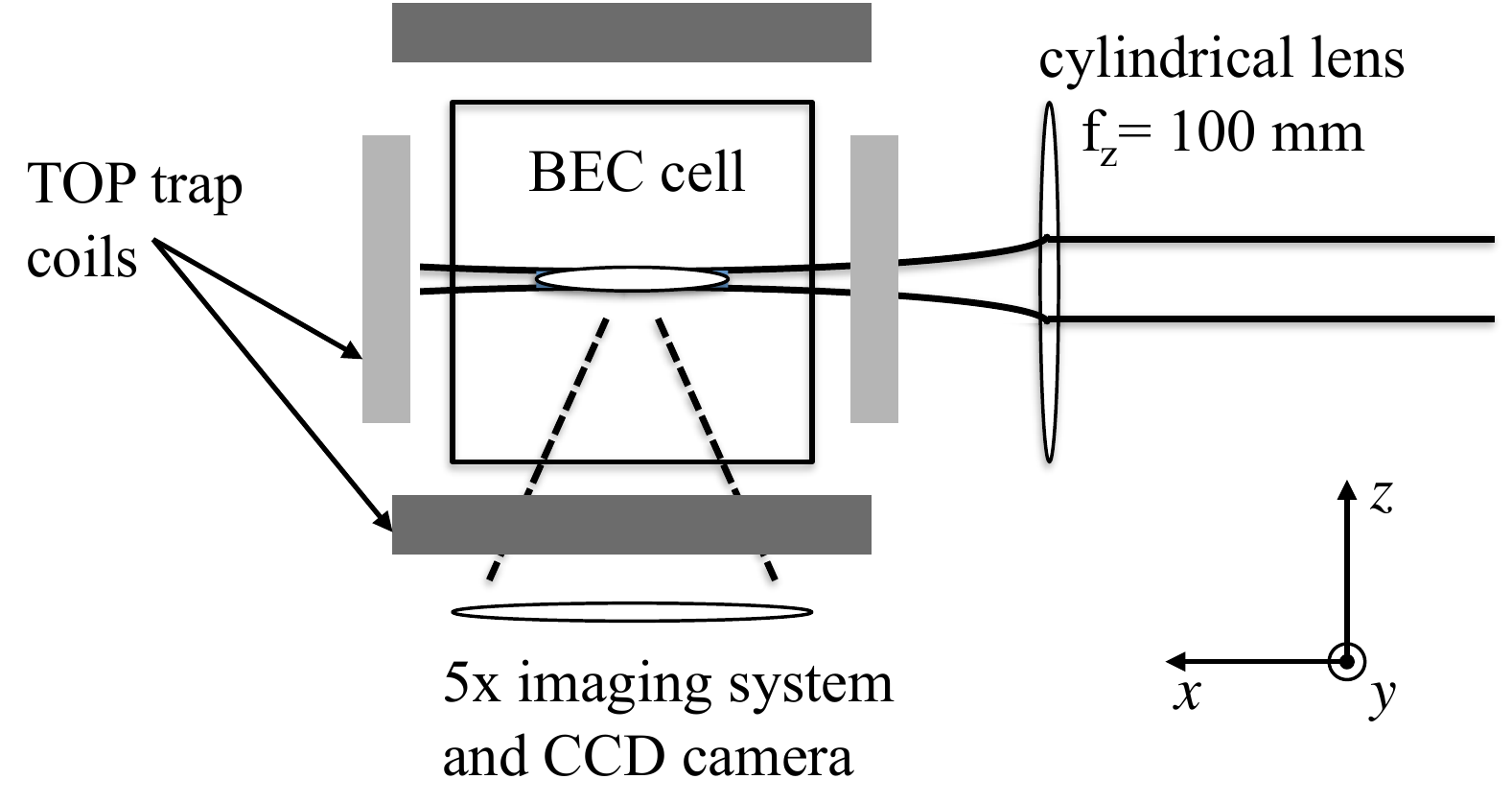}
\caption{Experimental configuration for the combined magnetic and optical harmonic trap. A cylindrical lens is used to focus a collimated beam of red-detuned laser light into a sheet with a vertical $1/e^2$ radius of $23 ~\mu$m at the location of the BEC. The coils for the DC component of the TOP trap, represented as dark gray bars, are located above and below the BEC cell.  The coils for the AC component, represented as light gray bars, are located on either side of the cell normal to the $x$-axis, and the $y$-axis (not shown).  A lens below the cell collects light for the vertical imaging system.} \label{exp_fig1}
\end{center}
\end{figure}

\begin{figure}
\begin{center}
\includegraphics[width=.8\linewidth]{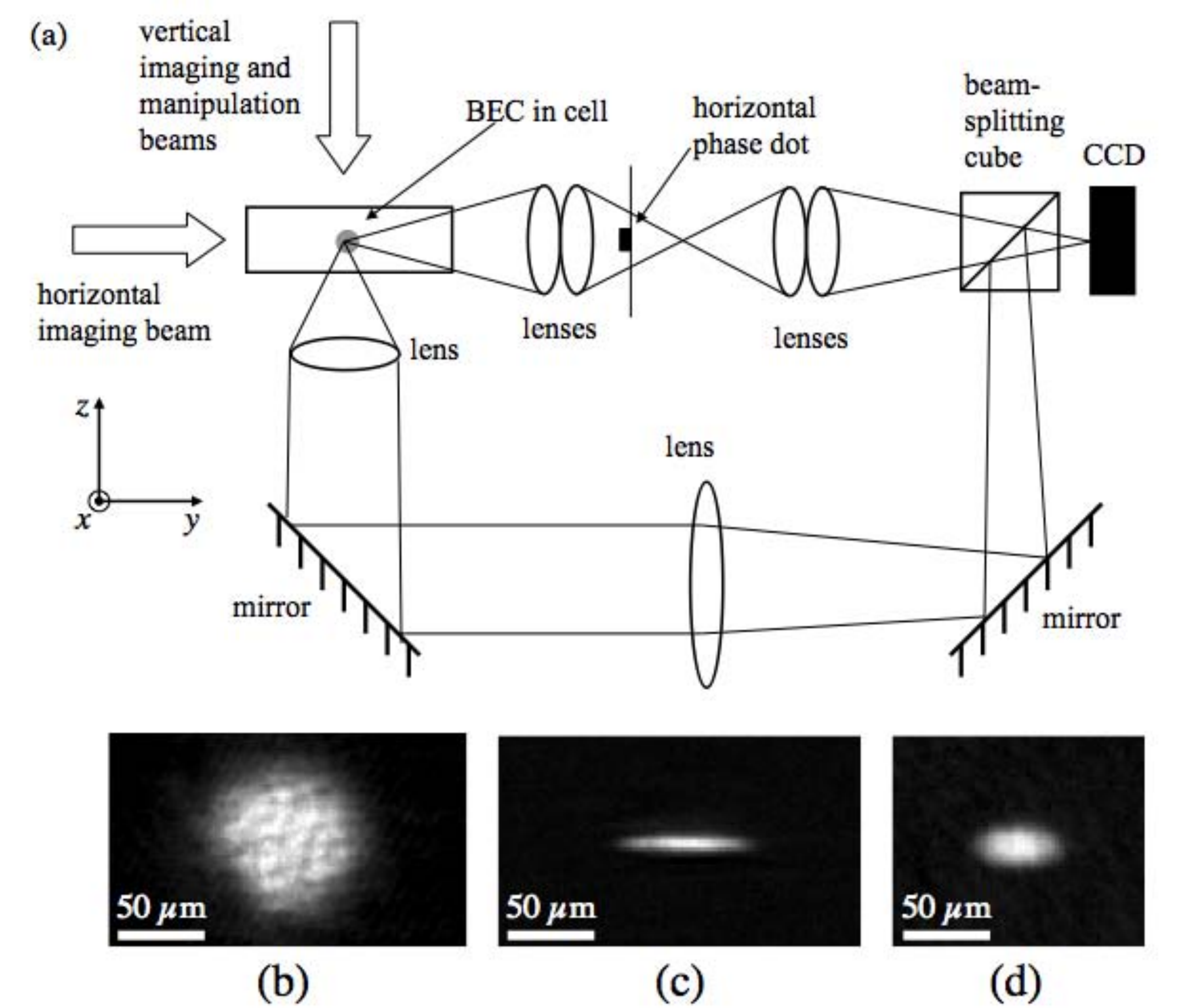}
\caption{BEC imaging. (a) illustration of the imaging layout. Both horizontal and vertical imaging systems have magnification M = 5. Side views of the BEC are obtained with phase contrast imaging along the horizontal imaging axis and top views are obtained with absorption imaging along the vertical.  (b) a vertically directed absorption image of the BEC in the combined magnetic and optical trap immediately after removal of the trapping potential but prior to any expansion.  (c) a corresponding \textit{in situ} horizontal phase-contrast image of the BEC in the combined magnetic and optical trap. (d)  \textit{in situ} horizontal phase-contrast image of the BEC in the purely magnetic TOP trap for comparison,  with an aspect ratio of Thomas-Fermi radii of $R_\mathrm{r}:R_\mathrm{z} =  2: 1$.} \label{exp_fig2}
\end{center}
\end{figure}

To obtain the highly oblate $^{87}$Rb BECs desired for our studies of vortex and 2DQT generation methods, we use a conventional magnetic time-averaged orbiting potential (TOP) trap \cite{petrich1995stc} overlaid with a 1090-nm red-detuned laser light sheet propagating along the $x$-axis as illustrated in \fref{exp_fig1} and \fref{exp_fig2}(a).  A cylindrical lens focuses the laser light in the $z$ direction only, with a waist at the position of the BEC such that the laser has horizontal and vertical $1/e^2$ radii of $(w_{0\mathrm{y}}, w_{0\mathrm{z}}) = (2000~ \mu\mathrm{m},\, 23 ~\mu\mathrm{m})$.   This red-detuned optical potential provides the additional axial ($z$) confinement needed to flatten the BEC while having a minimal effect on the radial ($r$) confinement.

We begin our experiments with evaporative cooling of atoms in the $|F=1,\,\mathrm{m}_{\mathrm{F} }= -1\rangle$ hyperfine state of $^{87}$Rb in the purely magnetic TOP trap. Prior to the final evaporation stage, we ramp on the trapping laser to a power of $\sim 1$ W over 4 s, and then cool the system below the BEC transition temperature $T_c \sim 100$ nK in the combined magnetic and optical potential.  The potential well that confines the BEC is an axially symmetric harmonic trap with radial and axial trap frequencies of  $(\omega_{\mathrm{r}}$, $\omega_{\mathrm{z}}) \sim 2\pi \times (8,90) ~\mathrm{Hz}$.  

The BECs formed in the combined magnetic and optical potential have up to $N \sim 2 \times 10^6$ atoms with a system temperature of $T \sim 50$ nK and Thomas-Fermi radii of $(R_{\mathrm{r}}, R_{\mathrm{z}}) \sim (52, 5) ~\mu\mathrm{m}$.  The chemical potential is $\mu_0 \sim 8\hbar\omega_\mathrm{z}$, and our BECs are consequently far from the quasi-2D limit and well into the 3D regime.   Nevertheless, the $R_\mathrm{r}:R_\mathrm{z} \sim  11: 1$ aspect ratio generated by the tight axial confinement suppresses vortex bending and tilting, resulting in a system characterized by approximately 2D fluid dynamics \cite{Rooney11a}.  Figures~\ref{exp_fig2}(b) and \ref{exp_fig2}(c) show images of a condensate held in the combined magnetic and optical trap.  For comparison, \fref{exp_fig2}(d) is an \textit{in situ} image of the BEC in the purely magnetic TOP trap with an aspect ratio of  $R_\mathrm{r}:R_\mathrm{z} =  2: 1$. Side views of the BEC are obtained with phase contrast imaging along the horizontal  ($y$) imaging axis and views along the vertical ($z$) direction are typically obtained with absorption imaging.

Vortex cores in our highly oblate BECs have a diameter on the order of the healing length,  $\xi \sim 0.4\, \mu$m for our system.  This is well below the resolution of our imaging system and it is currently beyond our ability to detect vortices with \emph{in situ} images.    In order to optically resolve the individual vortex cores of highly oblate BECs, we first turn off the TOP trap and allow the BEC to expand for 10 ms in the optical potential.  The optical trapping potential is then removed, and the BEC freely expands for an additional $40 ~\mathrm{ms}$.  During the entire expansion procedure, a separate magnetic field is applied with a gradient along the $z$ direction that cancels the downward gravitational force on the cloud.  \Fref{vortexfig} gives a representative absorption image after expansion, demonstrating that vortex cores are indeed resolved. In the remainder of this article, all images that show vortices are absorption images taken with the vertical imaging system after a total of $50$ ms of expansion.

\begin{figure}
\begin{center}
\includegraphics[width=.8\linewidth]{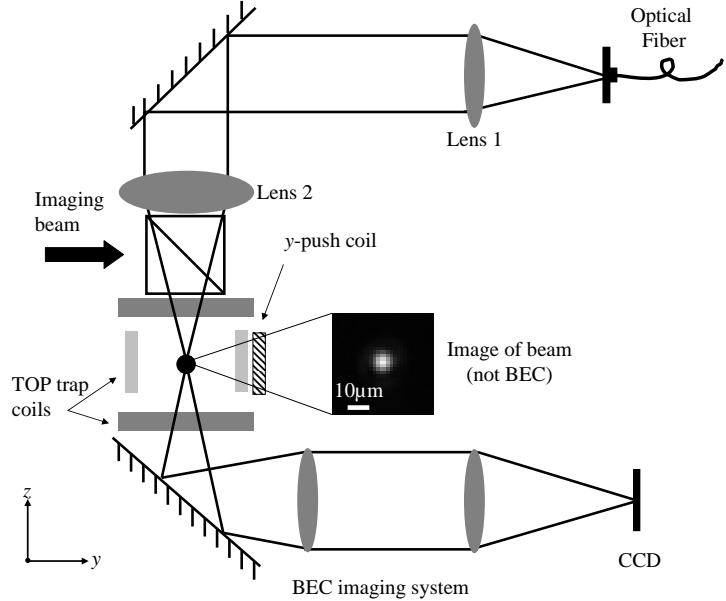}
\caption{Optical layout for the focused blue-detuned Gaussian beam directed axially through the BEC.  An image of the laser beam can be acquired using the vertical imaging system; an example image is shown.  The $y$ push coil (and a similar coil along $x$) can be used to manipulate the position of the center of the magnetic trap.} \label{exp_fig3}
\end{center}
\end{figure}

\begin{figure}[t]
\begin{center}
\includegraphics[width=4.5in]{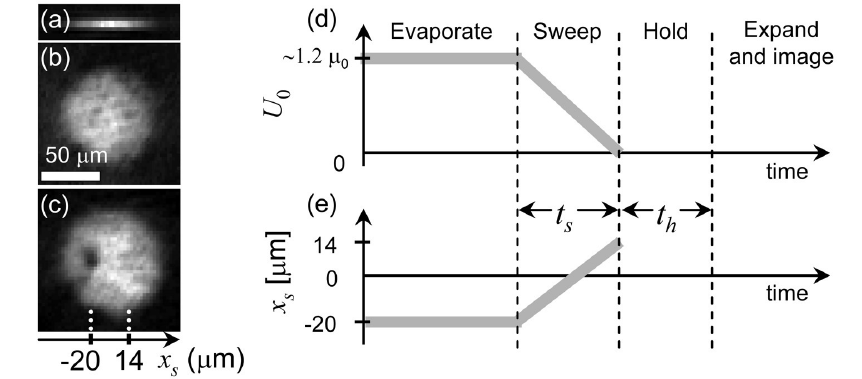}
\caption{The BEC initial state and experimental sequence.
(a) Side-view phase-contrast image and (b) axial absorption
image of a BEC in the highly oblate trap in the absence of the
obstacle. Lighter shades indicate higher column densities integrated
along the line of sight. Our axial and radial trapping frequencies are measured to be $\omega_z = 2 \pi \times 90$ Hz and $\omega_r = 2 \pi \times 8$ Hz, respectively. (c) BEC initial state with the obstacle located at $x_s = -20 \sim\mu$m relative to the BEC center. (d),(e) The maximum repulsive potential energy of the obstacle is $U_0 \approx 1.2\mu_0$ (where $\mu_0 \sim 8\hbar\omega_z$ is the BEC chemical potential) and is
held constant during evaporative cooling. It is ramped down
linearly as the trap translates; relative to the trap center, the beam
moves from position $x_s = -20 \sim\mu$m to $x_s = 14 \sim\mu$m over a
variable sweep time $t_s$. The BEC is then held in the harmonic
trap for a variable time $t_h$ prior to expansion and imaging.
Figure and caption from Ref.~[46], T. W. Neely,  E. C. Samson, A. S. Bradley, M. J. Davis, and B. P. Anderson, Phys. Rev. Lett.
\textbf{104}, 160401 (2010).  Copyright 2010 by the American Physical Society. }
\label{dipoles_fig1}
\end{center}
\end{figure}

In many of the experiments described in the following sections, we make use of an additional, focused blue-detuned Gaussian laser beam with a variable power and  $1/e^2$ beam radius, directed axially through the BEC as illustrated in \Fref{exp_fig3}.  This beam can be used to perturb and excite condensates through a variety of methods, which are described below.

We now turn to various experiments and tests of experimental methods, including relevant reviews of previous work in our lab on driving vortices into highly oblate BECs, and discussion of previously unpublished results on generation of 2DQT in highly oblate BECs.

\section{Nucleation of Vortex Dipoles and Vortex Clusters}\label{KW_sec4}

As described in Neely \textit{et al} \cite{Nee2010.PRL104.160401}, our group observed the deterministic formation and dynamics of vortex dipoles in a BEC by slightly stirring the BEC with a laser beam.  In this experiment, vortices were nucleated as the BEC was pushed around and past a blue-detuned laser beam that served as an impenetrable obstacle.  In the frame of the moving BEC, the beam was swept through the middle of the BEC, and the vortex dipoles formed in the wake of the laser beam.  A vortex dipole consists of a vortex and an antivortex -- two vortices of opposite circulation -- and tends to exist and move as a single excitation with linear momentum as long as the two vortices remain far from other vortices or the boundaries of the superfluid.  A vortex dipole can be generated from sound, hence a moving obstacle can excite a dipole in a compressible superfluid.  Similarly the vortices of a dipole can recombine, annihilating one another and generating a pulse of acoustic energy; see Ref.~\cite{Bradley2012} for a recent discussion of the role and consequences of compressibility in 2DQT.

\subsection{Overview of experiment}

In this experiment, the relative motion between the BEC and the blue-detuned laser beam occurred with a constant velocity, as illustrated in \fref{dipoles_fig1}.  BECs were created in a highly oblate harmonic trap as described above in \sref{KW_sec3} and shown in Figs.~\ref{dipoles_fig1}(a) and \ref{dipoles_fig1}(b). The focused blue-detuned Gaussian laser beam with a $1/e^2$ radius of $10~\mu$m was directed axially through the BEC 20 $\mu$m to the left of the BEC center as shown in \fref{dipoles_fig1}(c). A magnetic bias field was used to translate the harmonic trap minimum horizontally along the $x$-axis at a constant velocity until the laser beam was located 14 $\mu$m to the right of the center of the BEC. At the same time the intensity of the beam was ramped linearly from a maximum of $U_0 \sim 1.2\mu_0$ down to $U_0 = 0$ as shown in Figs.~\ref{dipoles_fig1}(d) and \ref{dipoles_fig1}(e). Here $\mu_0 \sim 8\hbar\omega_\mathrm{z}$ is the chemical potential of the BEC in the highly oblate harmonic trap. At the end of the sweep, the BEC was held in the harmonic trap for varying hold times $t_\mathrm{h}$ before expanding and imaging.

\begin{figure}[ht]
\begin{center}
\includegraphics[width=.9\linewidth]{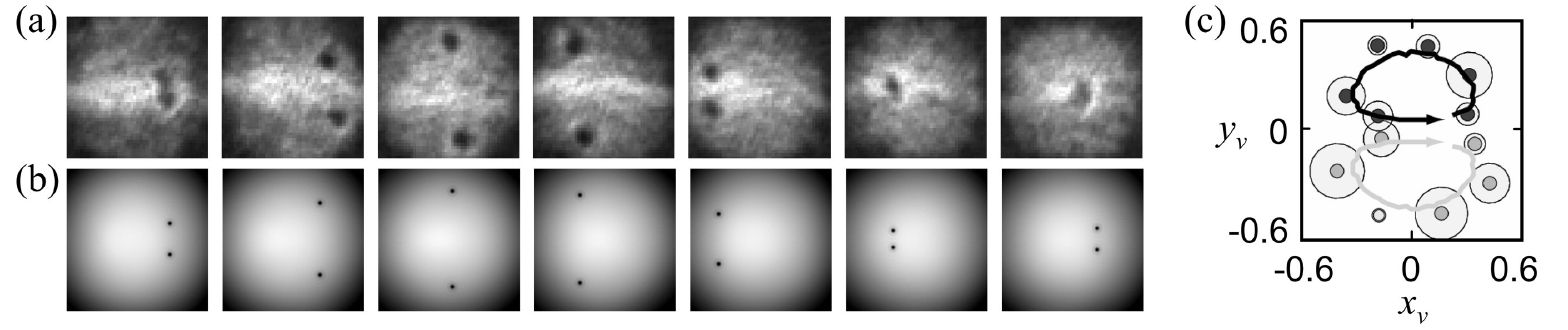}
\caption{Images showing the first orbit of vortex dipole dynamics. (a) Back-to-back expansion images from the experiment with 200 ms of successive hold time between the 180-$\mu$m-square images.  (b) 62-$\mu$m-square images from numerical calculations of the GPE obtained for conditions similar to the data of sequence (a), but for a temperature of $T = 0$ (i.e., no damping was used).  The orbital period is $\sim$ 1.2 s, and the apparent vortex core size is smaller in the simulations because we show in-trap numerical data. (c) Black and dark gray small circles show average positions of the vortices $x_v$ and $y_v$ (as fractions of the BEC radius) of each of the two vortices from 5 sequences of experimental data identical to that of sequence (a). The larger circle around each average position point represents the standard deviation of the vortex positions at that specific hold time, and is calculated from the 5 images obtained at that time step. A continuous dipole trajectory from sequence (b) is rescaled to the Thomas-Fermi radius of the expanded experimental images, and superimposed as solid lines on the experimental data; no further adjustments are made for this comparison.  Figure and caption adapted from Ref.~[46],  T. W. Neely, E. C. Samson, A. S. Bradley, M. J. Davis, and B. P. Anderson, Phys. Rev. Lett. \textbf{104}, 160401 (2010).  Copyright 2010 by the American Physical Society.}\label{dipoles_fig2}
\end{center}
\end{figure}

Neely \textit{et al} found a  critical velocity necessary for observing vortex dipole nucleation of 170-190 $\mu$m/s. The repeatability and coherence of this process allowed the dynamics of the vortex dipole to be mapped using images of BECs from multiple experimental runs, with each image taken at increasing values of $t_\mathrm{h}$. As shown in \fref{dipoles_fig2}, the vortex dipole orbited the BEC with a period of $\sim 1.2$ s.  GPE simulations of the procedure provided visual and quantitative comparisons with the experimentally observed vortex dynamics; see \fref{dipoles_fig2}(c).  Hall's group at Amherst College has also observed a variety of vortex dipole dynamics, using vortices spontaneously trapped in a BEC during the phase transition \cite{Weiler2008} and a detection method that permits multiple imaging of vortex positions in a single BEC \cite{Fre2010.Sci329.1182,Mid2011.PRA84.011605}.

\begin{figure}[ht]
\begin{center}
\includegraphics[width=.3\linewidth]{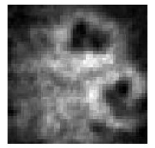}
\caption{An experimental image of a triply charged vortex dipole in a ballistically expanded BEC.   Adapted from Ref.~[46].}\label{dipole_clusters}
\end{center}
\end{figure}

Sweeping the BEC past the barrier at velocities much faster than the first critical velocity generated multiply quantized vortex dipoles. At a coarse scale, multiply charged dipoles behave like pairs of highly charged vortices of opposite circulation. The loci of vorticity orbits the condensate just as in the case of a singly charged dipole, albeit with a shorter period indicative of faster fluid flow.  A doubly charged vortex dipole, for example, orbited with a period of $\sim 0.8$ s.  As shown in \fref{dipole_clusters}, at finer scales these loci of vorticity are actually aggregates of singly quantized vortices of the same circulation. Both singly and multiply charged dipoles exhibited lifetimes up to many seconds indicating that vortex-antivortex recombination may be suppressed and that vortex dipoles may be meta-stable in highly oblate BECs.

\subsection{Implications for the experimental study of 2DQT}

The generation of acoustic energy and the nucleation of vortex dipoles due to a moving obstacle are associated with injection of kinetic energy (both compressible and incompressible) into the condensate.   The nucleation of vortices is also associated with the injection of enstrophy into the fluid, with the net enstrophy being proportional to the number of vortices \cite{Numasato10b,Bradley2012}.  We may view this basic laser sweep mechanism as a possible tool for the generation of 2DQT in a BEC, however in the experiment described above, the deterministic dynamics of the vortex dipoles indicate that a state of 2DQT was not observed.  The length scale associated with incompressible energy and enstrophy injection is most readily interpreted as the spatial separation between the vortices of the dipole at the instant of dipole nucleation; see Ref.~\cite{Bradley2012} for further discussion of this point.  With continuous laser stirring within the BEC, one may envision the generation of a 2DQT state, and indeed such a state has been observed as described in \sref{KW_sec6_3}.   See also the relevant simulations of Refs.~\cite{Tsubota2012} and \cite{Fujimoto2011} that involve an obstacle repeatedly sweeping through a BEC.

Whether or not a BEC with many vortices and antivortices would generally show traits associated with an inverse energy cascade or a direct energy cascade may strongly depend on specific experimental circumstances.  In particular, vortex-antivortex recombination is one mechanism for dissipation of enstrophy in 2D fluid flow and might inhibit the formation of an inverse energy cascade, especially for decaying 2DQT \cite{Numasato10b}.  The balance between vortex generation and vortex annihilation may influence observed characteristics of turbulence, such as energy spectra, kinetic energy flux, and aggregation of vorticity.  
 
Finally, the vortex aggregates observed in this experiment are directly generated by the relative motion between the laser obstacle and the BEC, and we do not interpret their deterministic formation as a consequence of 2D turbulence as proposed by Onsager \cite{Ons1949.NC6s2.279}.  Nevertheless, their long lifetimes (at least one orbital period) and steady alignment with the tight ($z$) trapping axis may be taken as an indication that vortex aggregates in a highly oblate BEC can be long-lived, and vortex-antivortex annihilation rates may be slow enough to reach regimes of constant or increasing net enstrophy.  In this sense, the observation of vortex aggregates in the experiment of Neely \textit{et al} gives hope for future experimental observation of vortex aggregates in 2DQT BEC experiments.


\section{Generating Turbulent States by Modulating the Magnetic Trapping Potential}\label{KW_sec5}

Our TOP trap combines a DC quadrupole magnetic field having variable axial gradient $B^{\prime}_\mathrm{z}$ with an AC bias field of magnitude B$_0$ that rotates in the $x$-$y$ plane with frequency $\omega_\mathrm{TOP} = 2\pi \times 4$ kHz.  The time-dependent bias field with $x$ and $y$ components
\begin{eqnarray}
B_\mathrm{x} = B_0 \cos (\omega _\mathrm{TOP} t)\label{eqn6} \\
B_\mathrm{y} = B_0 \sin (\omega _\mathrm{TOP} t)\label{eqn7} 
\end{eqnarray}
results in the minimum of the quadrupole field orbiting in a circle within the $x$-$y$ plane.
For an atom with magnetic dipole moment $\mu$ and mass $m$, the time average of the combined magnetic fields results in a harmonic potential  with trap frequencies determined by
\begin{eqnarray}
\omega_\mathrm{x} = \omega_\mathrm{y} = \Bigg[ \frac{\mu {B_\mathrm{z}^\prime}^2}{8 m B_0} (1 + \eta^2) \sqrt{1-\eta^2} \Bigg] ^{1/2} \\
\omega_\mathrm{z} = \Bigg[ \frac{\mu {B_\mathrm{z}^\prime}^2}{m B_0}  (1-\eta^2)^{3/2} \Bigg] ^{1/2}
\end{eqnarray}
where the factors involving $\eta = mg / \mu B_\mathrm{z}^\prime$ account for the vertical sag due to gravity (with acceleration $g$) of the TOP trap minimum from that of a static quadrupole field.  Such traps are now in common use; see Ref.~\cite{Ensher1998a} for a detailed description of TOP traps.  Modulations of the magnetic bias field and hence the shape of the trapping potential can excite various modes in the BEC.   For example, by deforming the trapping potential and rotating the spatial deformation axis, surface waves can be excited that decay to numerous vortices \cite{Abo2001.Sci292.5516,Mad2000.PRL84.806,Hod2001.PRL88.010405}.  In this section, we describe the vortex distributions and their lifetimes that result when highly oblate BECs are subjected to modulations of the magnetic component of the trapping field.


\subsection{Symmetric modulation: harmonic trap}\label{KW_sec5_1}

In this experiment, we modulated the amplitude of the magnetic bias field $B_0$ in Eqs. (\ref{eqn6}) and (\ref{eqn7}) by adding a small sinusoidal modulation to $B_0$ of the form
\begin{equation}
B_0(t) = B_0 + B_\mathrm{mod} \sin (\omega_\mathrm{mod} t).  
\end {equation}
Here $B_\mathrm{mod}$ is the amplitude of the bias field modulation and $\omega_\mathrm{mod}$ is the frequency of the modulation. This results in a small sinusoidal oscillation of the radial trap frequency $\omega_\mathrm{r}$ about the static value, but the trap remains axially symmetric. The axial trapping frequency $\omega_\mathrm{z}$ is primarily determined by the optical potential and therefore remains approximately constant.  Our aim in this experiment was to determine if a weak symmetric modulation would drive vorticity into the BEC such that there was approximately no net angular momentum transfer, i.e., such that if any vortices were generated, approximately equal numbers of vortices of both circulations would be created.  

In our procedure, we began with a highly oblate BEC in the combined magnetic and optical harmonic trap. We then modulated the amplitude of the rotating magnetic bias field at frequency $\omega_\mathrm{mod} = 2\pi \times 9 ~\mathrm{Hz}$ and amplitude $B_\mathrm{mod} \sim 0.05\, B_0$ for varying modulation time $t_\mathrm{mod}$. After the modulation we held the BEC in the static harmonic trap for varying hold times $t_\mathrm{h}$, then expanded and imaged, looking for the presence of vortices.

\begin{figure}
\begin{center}
\includegraphics[width=.8\linewidth]{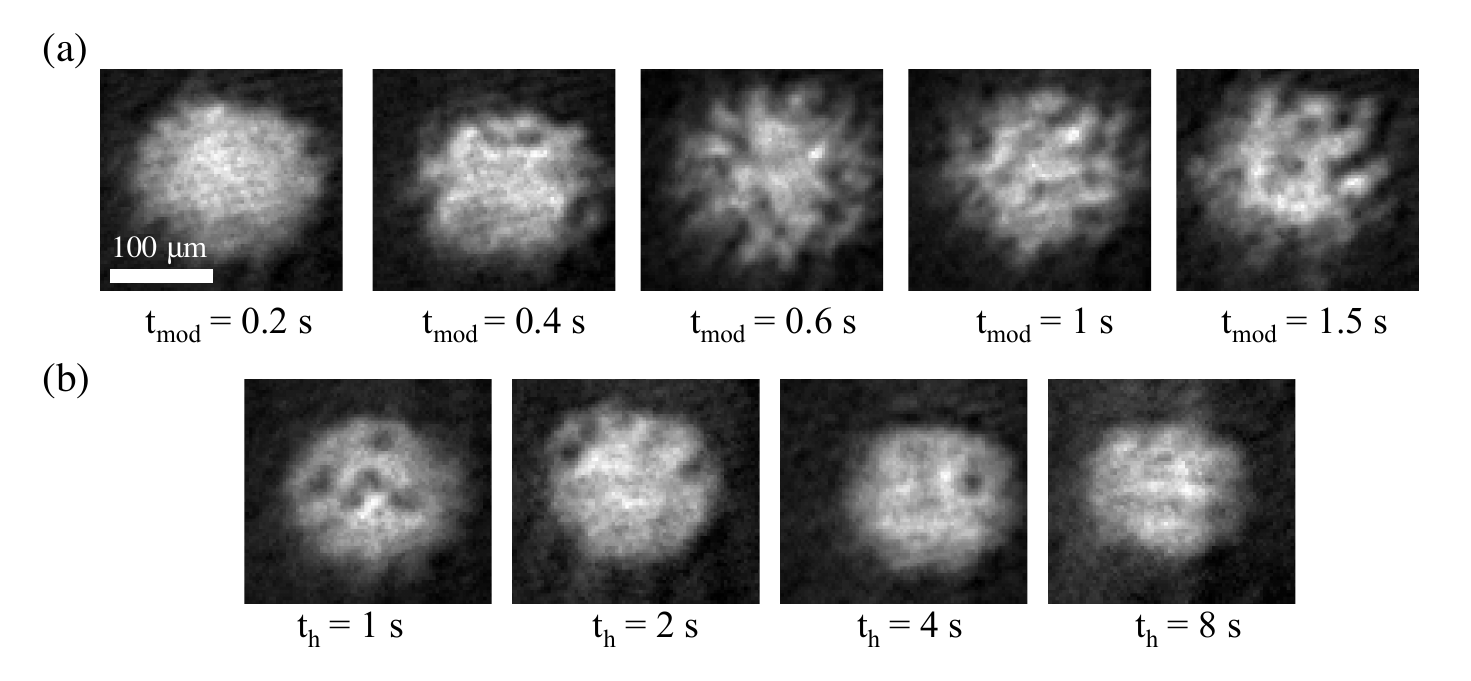}
\caption{Vortex distributions after varying modulation and hold times. (a) images taken at increasing modulation times from $t_\mathrm{mod}$ = 0.2 s to 1.5 s. Vortex cores start to appear at $t_\mathrm{mod}$ = 0.4 s and and by $t_\mathrm{mod}$ = 0.6 s we see a disordered vortex distribution. (b) images taken for increasing hold times after modulating for  $t_\mathrm{mod}$ = 1.5 s. } \label{modB_fig1}
\end{center}
\end{figure}

\Fref{modB_fig1}(a) shows images taken at increasing modulation times from $t_\mathrm{mod}$ = 0.2 s to 1.5 s. Vortex cores started to appear at $t_\mathrm{mod}$ = 0.4 s and by $t_\mathrm{mod}$ = 0.6 s we observed disordered vortex distributions. Vortex cores appear to be generated at the outer edge of the condensate and make their way into the center. In the next step of the experiment, we modulated the trap as described for $t_\mathrm{mod}$= 1.5 s  before removing the modulation and returning the BEC to the unmodulated harmonic trap.  We added increasing post-modulation hold times $t_h$ and then expanded and imaged the BEC. As shown in \fref{modB_fig1}(b) the vortex distribution became better resolved with increasing $t_h$ and decayed to states with only a few cores remaining after $t_\mathrm{h}$ = 2 s.  By $t_\mathrm{h}$ = 8 s no vortex cores remained. The condensate was completely destroyed for $t_\mathrm{mod}$ = 10 s.

\subsection{Symmetric modulation: annular trap}

\begin{figure}[t]
\begin{center}
\includegraphics[width=.9\linewidth]{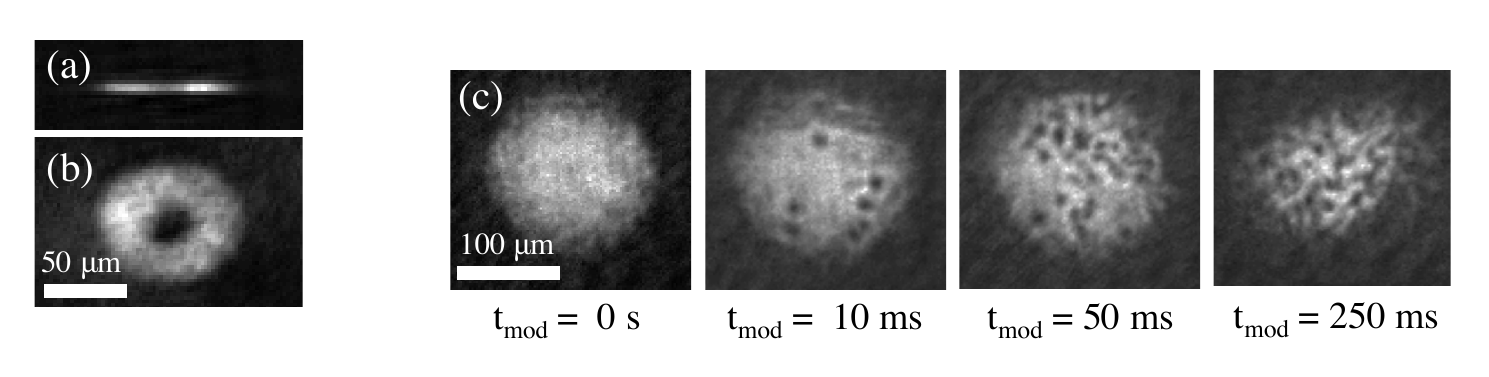}
\caption{Modulating the radial trapping frequency for a BEC in an annular trapping geometry. (a)  side view of an annular BEC imaged using \textit{in situ} phase contrast imaging. (b) top-down view of an annular BEC imaged using absorption imaging immediately after turning off the trapping potential. For these two images, the blue-detuned beam provides a barrier of height well above the chemical potential, confining the atoms to an annular trap. The annular trap used for the sequence of images in (c) had a narrower, weaker central barrier. (c) images taken at increasing modulation times with no hold after modulation. The first image shows the expanded BEC for $t_\mathrm{mod}$ = 0 s to show that the hole created by the blue-detuned beam has completely filled between beam ramp down and imaging. For the annular trapping geometry vortices appear after much shorter modulation times than for the harmonic trapping case.  }  \label{modB_fig2}
\end{center}
\end{figure}

In a similar experiment, we modulated the trapping potential of a highly oblate annular trap.  To do this, we modulated the strength of the rotating bias field of the TOP trap in the same manner as described in \sref{KW_sec5_1}.  We added to the harmonic trap a focused blue-detuned Gaussian laser beam, directed axially through the BEC in the configuration shown in \fref{exp_fig3}. Example images of such a BEC are shown in Figs. \ref{modB_fig2}(a) and \ref{modB_fig2}(b). Prior to imaging, the blue-detuned laser barrier was ramped off slowly enough to avoid significantly perturbing the condensate, allowing the BEC to expand from a harmonic trap for imaging.  With $\omega_\mathrm{mod} = 2\pi \times 6.5 ~\mathrm{Hz}$ and $B_\mathrm{mod} \sim 0.1 B_0$, numerous vortices were quickly nucleated over just a few ms, a significant change from the harmonic trap. As shown in \fref{modB_fig2}(c) vortex cores were observed to be distributed throughout the BEC, with the first cores appearing at least an order of magnitude earlier than in the purely harmonic trap.  Compared to the results of \sref{KW_sec5_1}, many more cores were visible using this modulation technique in the annular BEC, and the cores have higher contrast.  We speculate that  the presence of the blue-detuned barrier acted as a vortex dipole nucleation site located within the BEC, and that the vortex dipoles quickly dissociated and formed a state of 2DQT.  Again, the modulation process was designed to be axially symmetric apart from experimental errors in beam positioning and trap roundness, and should have nominally added no net angular momentum to the condensate.

With additional modulation time, the BEC was driven into a state of high excitation in which individual vortex cores were not easily resolvable or countable.  Such a state is shown in the rightmost image of   \fref{modB_fig2}(c).  This process shows that too much modulation drives the BEC into a highly excited far-from-equilibrium state, a signature that it may be possible to identify a weaker modulation strength such that vortex nucleation rates and vortex decay rates due to annihilation and thermal damping at the system boundary would be balanced for a given temperature.  Finding this optimum balance is a goal of future research.


\subsection{Rotation of an elliptical magnetic trapping potential}\label{KW_sec5_2}

The straightforward experimental method of expansion imaging used in our experiments does not easily permit measurement of the circulation of vortices.  While the interferometric or dynamic methods mentioned above can be used for this purpose, these have not yet been applied to turbulent states of BECs.  In order to look for the clustering of vortex cores of like-circulation, an alternative is to examine statistics of vortex distributions containing only large numbers of cores of identical circulation.  One possible approach to generate such distributions is by rotating the trapping potential, as has been utilized in numerous experiments of BECs with different aspect ratios, such as Refs.~\cite{Abo2001.Sci292.5516,Mad2000.PRL84.806,Hod2001.PRL88.010405}; see also Ref.~\cite{Anderson2010} for an overview of numerous other experiments that utilized this technique.  Our approach follows such previous work, and is implemented by squeezing and rotating the magnetic trapping field as first implemented by Hodby \textit{et al}\cite{Hod2001.PRL88.010405}. 

\begin{figure}
\begin{center}
\includegraphics[width=.3\linewidth]{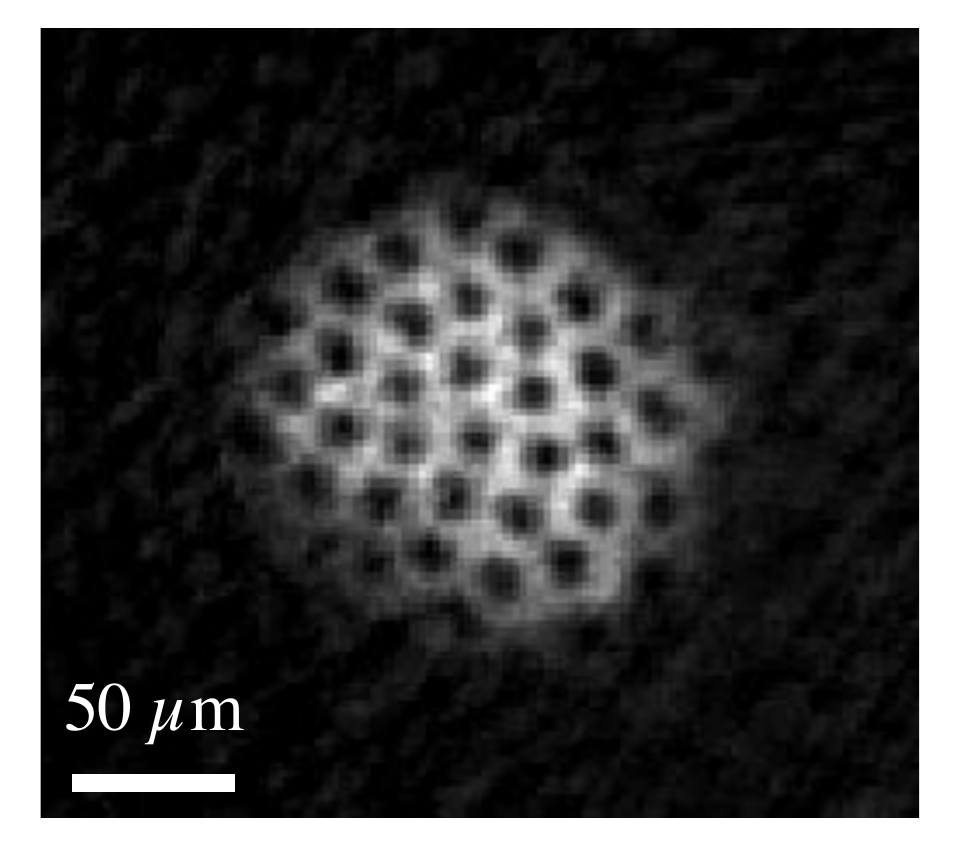}
\caption{An image of a slightly irregular vortex lattice in a rotating BEC released from our TOP trap.} 
\label{vortexlat}
\end{center}
\end{figure}

In order to rotate and add angular momentum to the atomic cloud it is necessary to break the symmetry of the trapping potential.  In our case this is done by adding an additional bias field to the one described by Eqs.~(\ref{eqn6}) and (\ref{eqn7}) so that now
\begin{eqnarray}
B_\mathrm{x} = B_0 \cos (\omega _\mathrm{TOP} t) + B_\epsilon \cos (\omega _\mathrm{mod} t)\\
B_\mathrm{y} = B_0 \sin (\omega _\mathrm{TOP} t) - B_\epsilon \sin (\omega _\mathrm{mod} t)
\end{eqnarray}
where $B_\epsilon \sim 0.1\,B_0$, and $\omega_\mathrm{mod}$ is the frequency of the modulating bias field. If $\omega_\mathrm{mod} = \omega_\mathrm{TOP}$ the bias field causes the time-averaged trap to have stationary ellipsoidal potential energy surfaces in the $x$-$y$ plane with the ratio of the minor and major axes determined by $B_\epsilon$ and $B_0$; for our case this ratio is  $\sim 0.8$. Increasing $B_\epsilon$ with respect to $B_0$ increases this ellipticity.  If $\omega_\mathrm{mod} \neq \omega_\mathrm{TOP}$ the ellipse rotates in the $x$-$y$ plane with frequency $\omega_\mathrm{s} = |\omega_\mathrm{mod} - \omega_\mathrm{TOP}| / 2$. In the TOP trap, choosing $\omega_\mathrm{s} \sim 0.7\omega_\mathrm{r}$ excites the quadrupole mode which decays to a collection of quantized vortex cores of the same circulation. The minimum energy configuration of the rotating BEC consists of a large regular lattice of vortices \cite{Abo2001.Sci292.5516}.   A representative vortex lattice formed by spinning the BEC in our purely magnetic TOP trap (i.e., without the red-detuned trapping potential) is shown in \fref{vortexlat}.

Our investigations of highly oblate BECs subjected to trap deformation and rotation started with a BEC in the combined magnetic and optical harmonic trap. We then applied the magnetic field ellipticity and spun in the highly oblate harmonic trap for time $t_\mathrm{s}$ at frequency $\omega_\mathrm{s}$ with ellipticity $B_\epsilon$. After spinning we returned to the symmetric harmonic trap and held for time $t_\mathrm{h}$ while the BEC shape deformations damped out.  Finally, we let the condensate expand and we imaged the cloud.   

\begin{figure}
\begin{center}
\includegraphics[width=.8\linewidth]{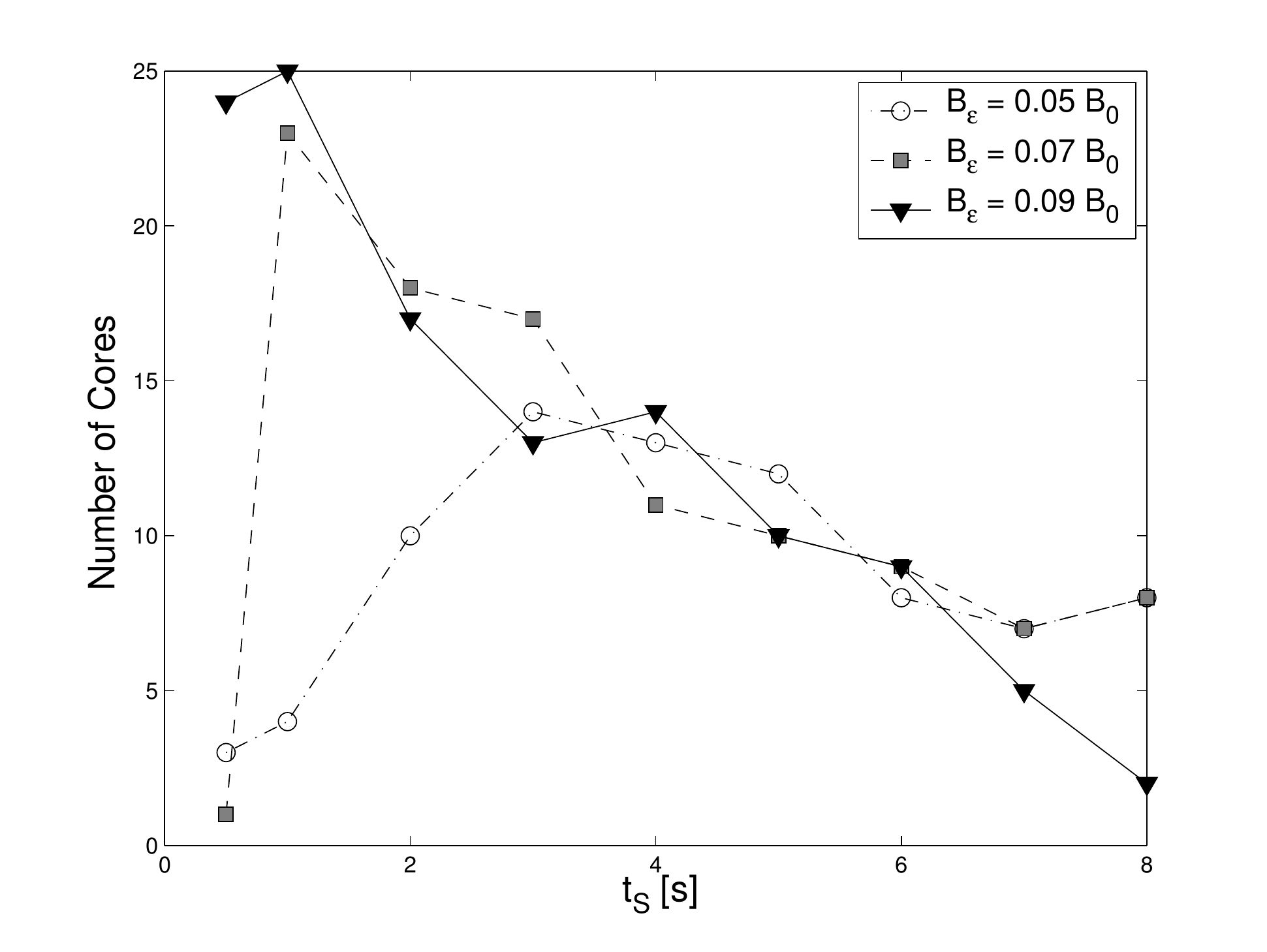}
\caption{Comparison of number of vortex cores versus spin time for different ellipticities. All data points were generated by spinning at $\omega_\mathrm{s} = 2\pi \times 6$ Hz for a variable spin time $t_{\mathrm{s}}$ and then holding for 2.5 s after the spin prior to expansion. Open circles represent a single set of data taken with $B_\epsilon = 0.05\, B_0$, gray squares represent $B_\epsilon = 0.07\, B_0$, and black triangles represent $B_\epsilon = 0.09\, B_0$.  Both the peak number of vortex cores and $t_{\mathrm{s}}$ corresponding to the peak number of vortex cores seem to be dependent on ellipticity but all three ellipticities show decreasing vortex number for longer spin times.} \label{spin_fig1}
\end{center}
\end{figure}

By varying spin frequencies, we found a resonance condition for generating large numbers of vortices at $\omega_\mathrm{s} = 2 \pi \times 6$ Hz. Here $\omega_\mathrm{s} \sim 0.75\omega_\mathrm{r}$ is consistent with previous observations of exciting the quadrupole mode in 3D harmonic traps at  $\omega_\mathrm{s} \sim 0.7\omega_\mathrm{r}$ \cite{Che2000.PRL85.2223}.  We measured the number of vortex cores versus spin time for different ellipticities at this resonance frequency. \Fref{spin_fig1} shows three sets of data corresponding to ellipticities of $B_\epsilon = 0.05\,B_0$, $0.07\,B_0$ and $0.09\,B_0$. All data points were generated by spinning for a variable spin time $t_{\mathrm{s}}$ and then holding for $t_{\mathrm{h}}$ = 2.5 s after the spin and before expansion. Both the peak number of vortex cores and $t_{\mathrm{s}}$ corresponding to the peak number of vortex cores appear to be dependent on ellipticity with peak vortex number occurring at a much shorter spin time for $B_\epsilon = 0.09\, B_0$ and $0.07\, B_0$ than for $B_\epsilon = 0.05\, B_0$. All three ellipticities show decreased vortex numbers for longer spin times on the order of $t_{\mathrm{s}}$= 8 s indicating that vortex cores may be annihilating or leaving the system before crystalizing into a lattice. 

\begin{figure}
\begin{center}
\includegraphics[width=.8\linewidth]{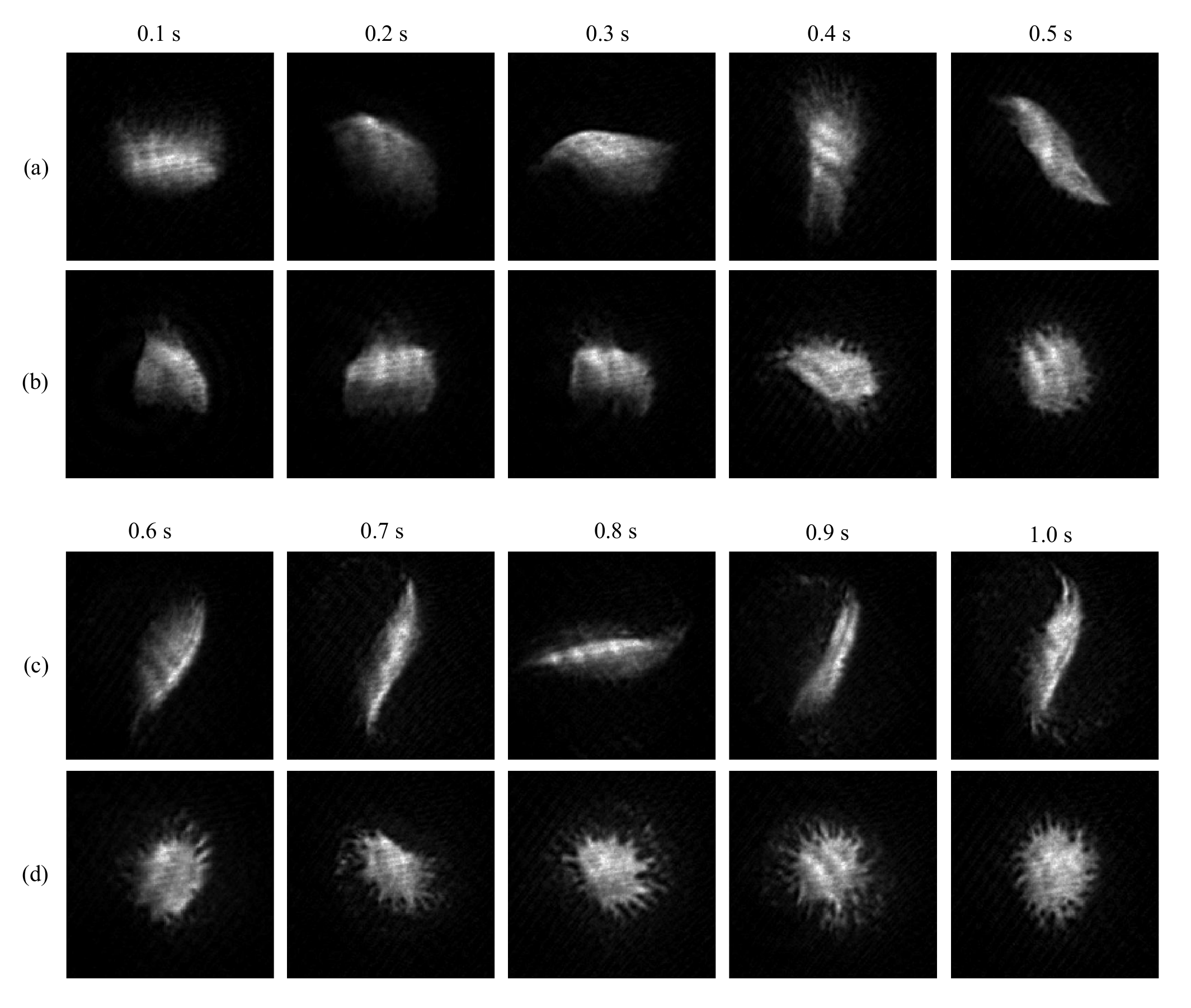}
\caption{350-$\mu$m-square expansion images of an oblate BEC for varying $t_\mathrm{s}$ (indicated above images) at increments of 100 ms. Spin frequency and ellipticity were held constant at $\omega_\mathrm{s} = 2\pi \times 6$ Hz and $B_\epsilon = 0.07 B_0$.  (a) and (c) BECs were expanded and imaged directly after spinning, $t_\mathrm{h}$ = 0.  (b) and (d) BECs were held for 400 ms after spinning, $t_\mathrm{h}$ = 400 ms, then expanded and imaged.} \label{spin_fig2}
\end{center}
\end{figure}

\Fref{spin_fig2} visually tracks the evolution of the condensate during spinning.  Figures~\ref{spin_fig2}(a) and \ref{spin_fig2}(c) show successive images at 100 ms intervals for $t_\mathrm{s}$ = 0.1 to 1.0 s. The BECs were imaged directly after spinning with no hold in the axially symmetric harmonic trap ($B_\epsilon = 0$) prior to expansion. Figures~\ref{spin_fig2}(b) and \ref{spin_fig2}(d) show successive images at 100 ms intervals for $t_\mathrm{s}$ = 0.1 to 1.0 s with an additional 400 ms hold time in the axially symmetric harmonic trap prior to expansion and imaging.  The hold time in the symmetric harmonic trap seems important for the nucleation of vortices as vortex cores appear after $t_{\mathrm{s}}$ = 400 ms and $t_{\mathrm{h}}$ = 400 ms, but do not appear until $t_{\mathrm{s}}$ = 1.0 s for the case of $t_{\mathrm{h}}$ = 0. 

\begin{figure}
\begin{center}
\includegraphics[width=.8\linewidth]{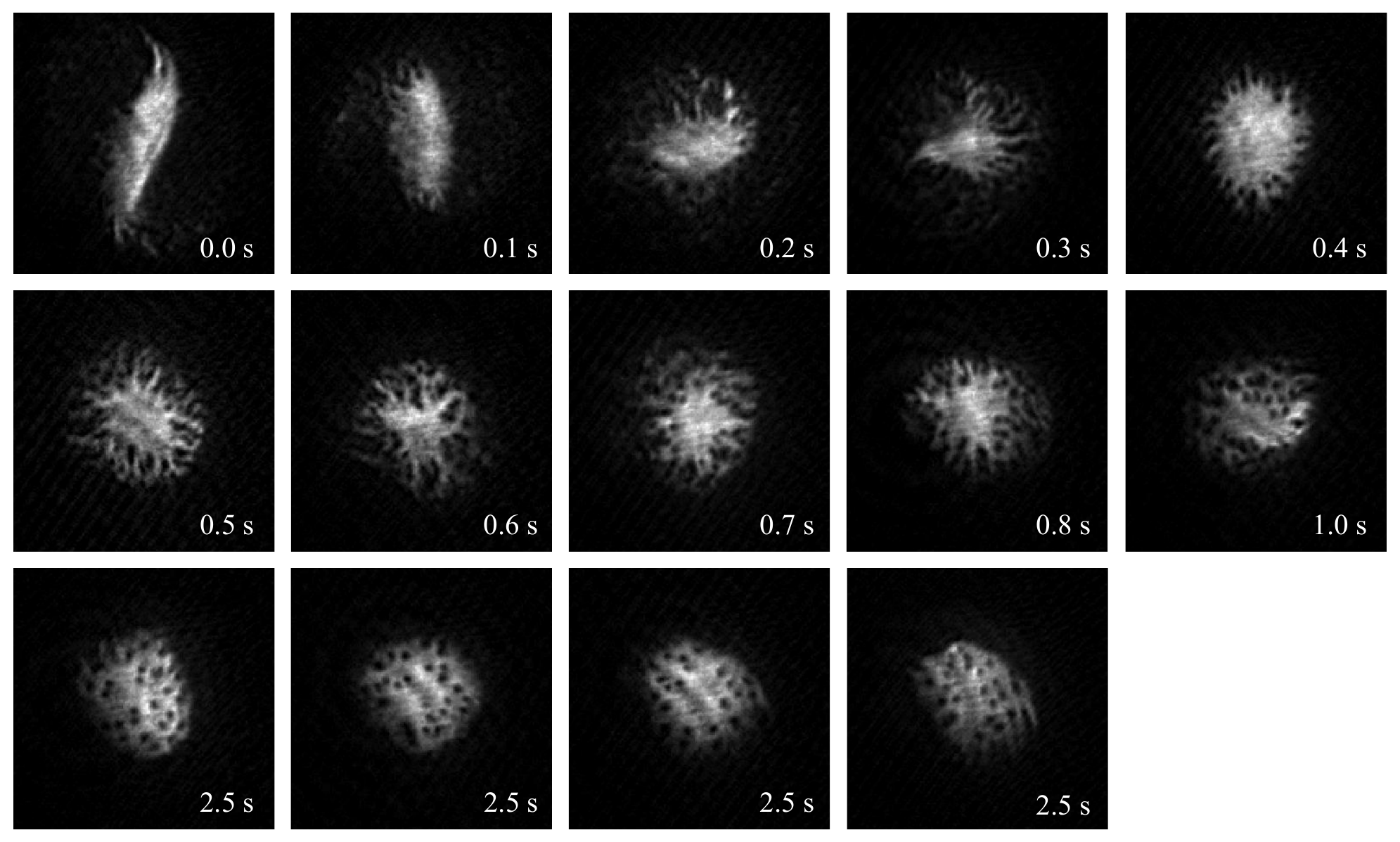}
\caption{350-$\mu$m-square expansion images of vortex distributions in the highly oblate BEC for varying $t_\mathrm{h}$.  Spin time, frequency, and ellipticity were held constant at $t_\mathrm{s}$ = 1.0 s, $\omega_\mathrm{s} = 2\pi \times 6$ Hz and $B_\epsilon = 0.07 B_0$.   Images were taken for variable $t_\mathrm{h}$  from 0 to 2.5 s. Multiple images for $t_\mathrm{h}=2.5$ s are shown to give a representative sample of the variation in vortex distribution.} \label{spin_fig3}
\end{center}
\end{figure}

\begin{figure}
\begin{center}
\includegraphics[width=.8\linewidth]{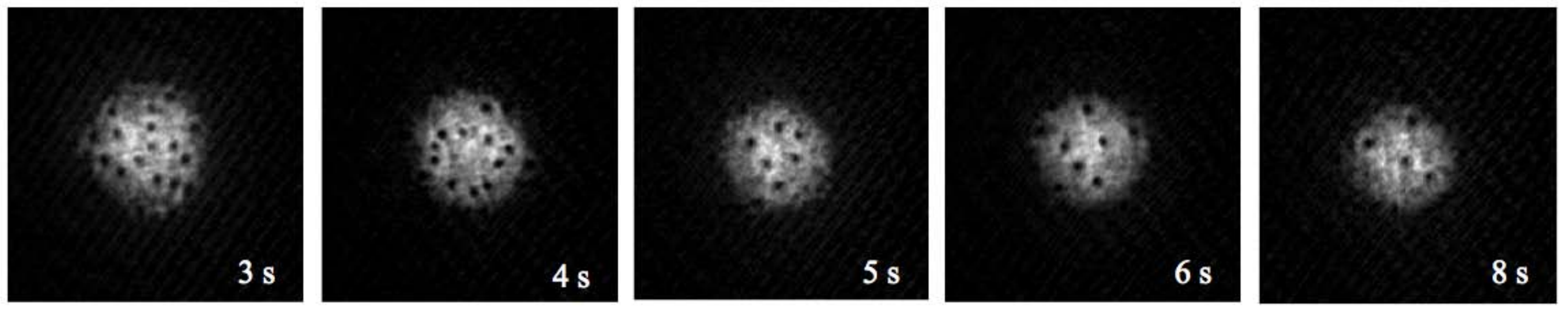}
\caption{350-$\mu$m-square expansion images of vortex distributions in the highly oblate BEC for the varying hold times $t_\mathrm{h}$ shown from 3 to 8 s.  By this time, BEC shape deformations have largely damped out.  Spin time, frequency, and ellipticity were held constant at $t_\mathrm{s}$ = 1.0 s, $\omega_\mathrm{s} = 2\pi \times 6$ Hz and $B_\epsilon = 0.07 B_0$.} \label{spin_fig4}
\end{center}
\end{figure}

Figures \ref{spin_fig3} and \ref{spin_fig4} show additional distributions of vortex cores nucleated by spinning and relaxing over subsequent hold times. Here $t_\mathrm{s} = 1.0 ~\mathrm{s}$ is held constant and the BECs are imaged after increasing values of $t_{\mathrm{h}}$. Vortices appear to be nucleated on the outer edge of the BEC. Presumably there is net angular momentum added to the condensate, so we expect the vortex cores to be predominately of the same circulation, at least once the system has relaxed to a meta-stable configuration. However, it may be the case that the net circulation is large, but that numerous vortices of the opposite circulation are also generated and present in these images, at least prior to the point at which a disordered distribution is uniformly distributed throughout the BEC.  If vortices are indeed of the same circulation in images such as the ones in the bottom row of \fref{spin_fig3}, then such states may be candidates for experimental measurement of vortex power-law distributions in a BEC, as has been analytically described for homogeneous BECs \cite{Bradley2012}.

As observed,  large, disordered vortex distributions can be generated by exciting collective modes of the BEC.  The shape deformations damp more quickly than the number of vortex cores, leaving open the possibility of finding parameters where studies of decaying 2DQT could be performed before the system spins down to a state with no vortices.  Spinning introduces a net angular momentum into the condensate and should result predominately in vortex cores of the same circulation, making this a possible system in which to observe vortex aggregates without time-resolved dynamics measurements.  Nevertheless, this vortex excitation technique does not appear to satisfy the particular goal of continuous forcing, although it appears to be a candidate for studies of decaying 2DQT. It is interesting to note that even for long hold times we find no evidence of a vortex lattice in the highly oblate BEC.  Thermalization times for a lattice in a highly oblate BEC may be beyond the lifetime of our BECs; see Ref.~\cite{Wright2008a} for further discussion of this issue.


\section{Generating Turbulent States with a Stationary Blue-Detuned Laser}\label{KW_sec6}

In the experiments described in this section,  we investigated the response of a BEC to time-dependent perturbations of the intensity of a focused blue-detuned laser beam that pierced the BEC.  In all cases in this section, the relative position between the beam and the BEC was stationary.  Our aim in these experiments was to locally excite the BEC as an empirical probe of the existence of thermal counterflow \cite{Vin1957.PRSLA242.493,Vin1958.PRSLA243.400} in a region where the BEC was locally depleted of atoms.  While vortices were observed in all methods examined, the mechanisms for vortex nucleation remain unclear and merit further experimental and numerical investigation.


\subsection{Short pulse of blue-detuned laser light}\label{KW_sec6_1}

\begin{figure}
\begin{center}
\includegraphics[width=.8\linewidth]{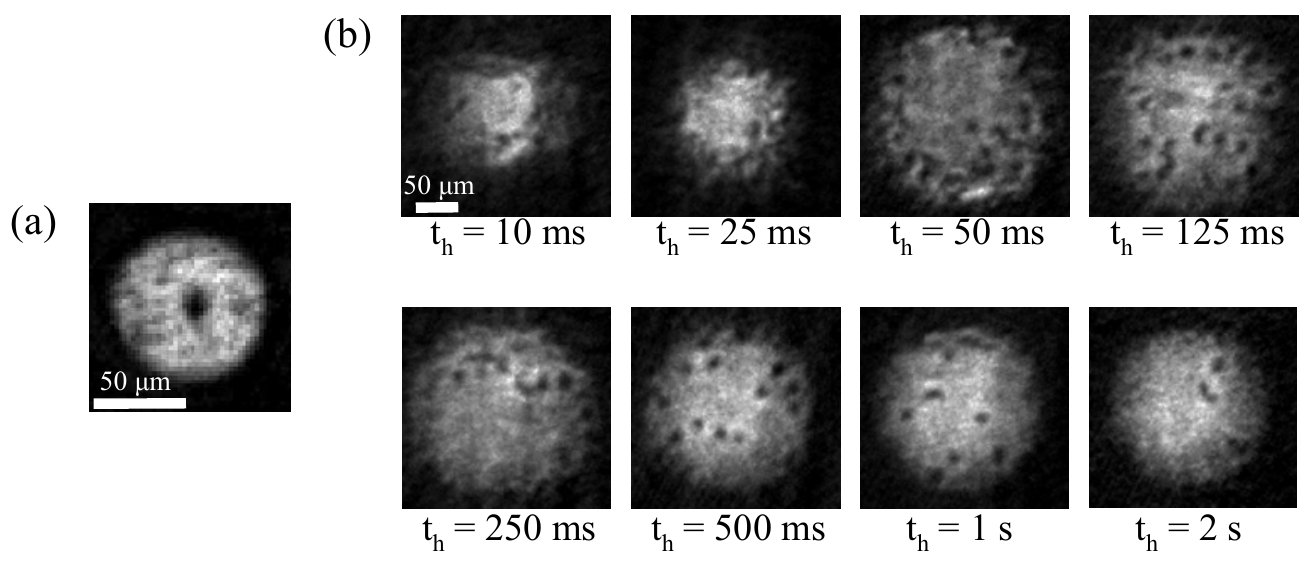}
\caption{Vortices generated by subjecting the BEC to a 7 ms pulse of blue-detuned laser light directed axially through the center of the condensate. (a)  \textit{in situ} absorption image of the BEC with the blue-detuned beam penetrating the center of the BEC. The beam intensity is chosen such that the beam potential is approximately equal to the condensate chemical potential.  (b) Images show successive hold times after the 7ms laser pulse.  } \label{blast_fig1}
\end{center}
\end{figure}

In this experiment we nucleated vortices by subjecting the BEC to a short pulse of blue-detuned laser light. We began by forming a highly oblate BEC in the purely harmonic trap. We then instantaneously turned on a blue-detuned Gaussian beam for a pulse time of 7 ms, short compared to the radial harmonic oscillator period of 125 ms. The focused blue-detuned Gaussian beam had a $1/e^2$ radius of 10 $\mu$m and was directed axially through the center of the condensate, as shown in \fref{exp_fig3}. The power in the blue-detuned beam was chosen such that the optical potential generated by the  beam was approximately equal to the chemical potential of the BEC. After the laser pulse we held the BEC in the harmonic trap for varying hold times $t_\mathrm{h}$ before expansion and imaging. \Fref{blast_fig1}(a) shows an \textit{in situ} absorption image of the BEC with the blue-detuned beam on, clearly penetrating the condensate. 

The laser pulse generated a shock wave that can be observed as the central high density region in the first image in \fref{blast_fig1}(b). Subsequent images show the condensate for longer hold times. Vortices appear to enter  the condensate from the outer boundary as the shock wave propagates. Vortices eventually leave the system, most likely through vortex-antivortex recombination and thermal damping.  Apart from the shock wave, this perturbation created vortices with little residual excitation of the BEC.  Hoefer \textit{et al}  \cite{Hoefer2006} performed experimental and numerical studies of the dispersive shockwaves generated when a BEC was subjected to a short pulse of laser light, however their experiments were performed with a prolate BEC and they do not report observations of vortex cores generated as a result of the laser pulse.


\subsection{Intensity modulation of a blue-detuned laser beam}\label{KW_sec6_2}

\begin{figure}[t]
\begin{center}
\includegraphics[width=.8\linewidth]{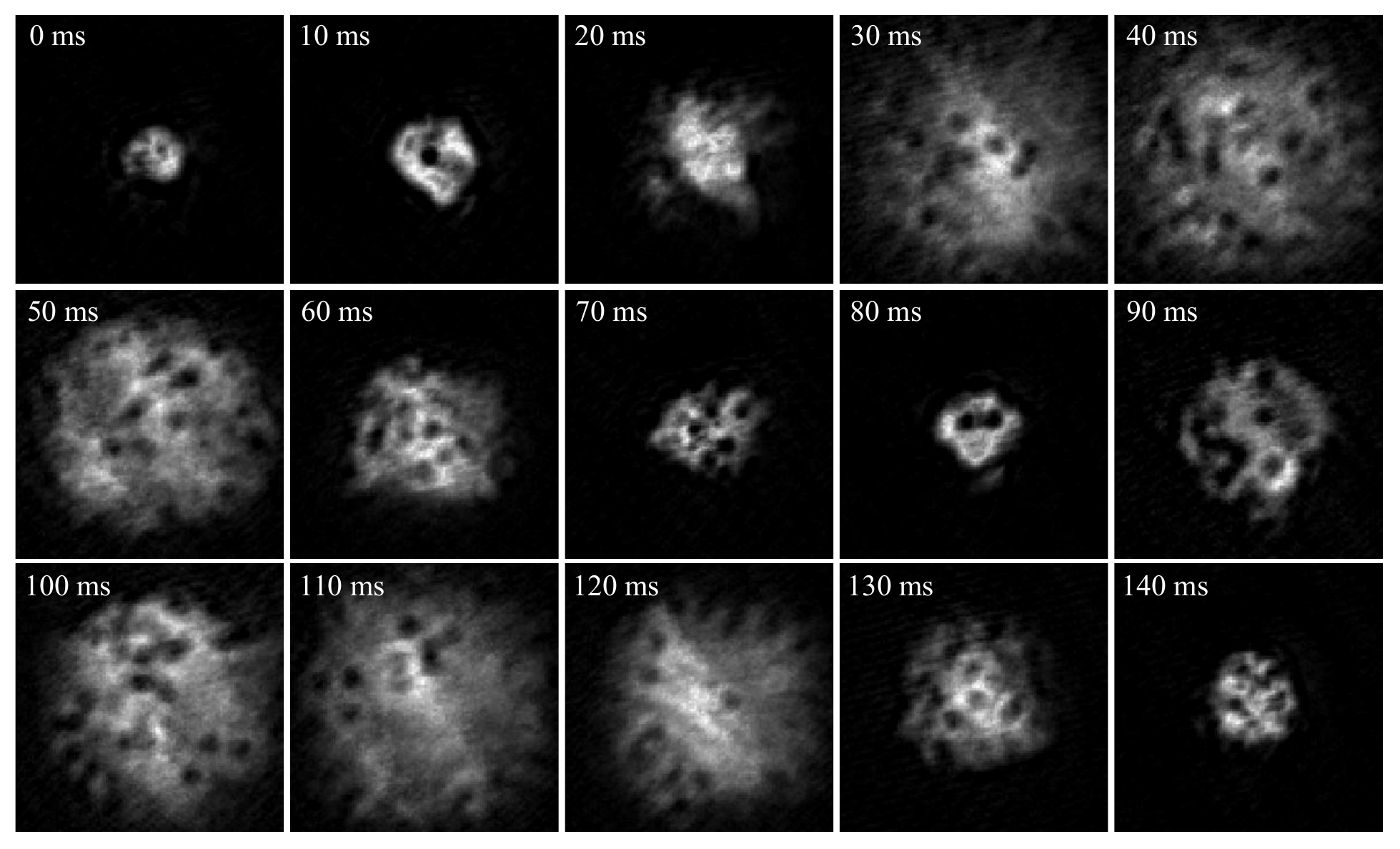}
\caption{250-$\mu$m-square absorption images acquired after the hold times $t_\mathrm{h}$ indicated. The condensate undergoes large breathing oscillations in the trap, and this oscillation leads to a periodic variation in expanded BEC radius between $\sim40\,\mu$m and $\sim180\,\mu$m. } \label{660mod_fig1}
\end{center}
\end{figure}

\begin{figure}[h]
\begin{center}
\includegraphics[width=.7\linewidth]{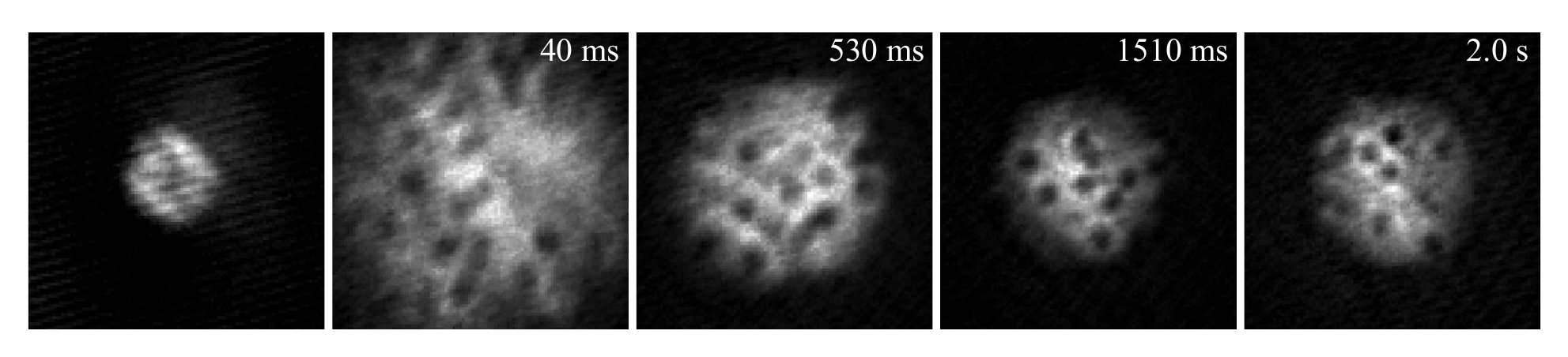}
\caption{250-$\mu$m-square absorption images acquired after the hold times $t_\mathrm{h}$ indicated. The first image is an \textit{in situ} image of the BEC in the harmonic trap with the blue-detuned beam partially penetrating the condensate; note that the optical potential strength is well below the BEC chemical potential.} \label{660mod_fig2}
\end{center}
\end{figure}

In this experiment we formed a highly oblate BEC in the purely harmonic trap, then turned on and sinusoidally modulated the intensity of a focused blue-detuned Gaussian beam for a varying time $t_\mathrm{mod}$. After the modulation we held the BEC in the purely harmonic trap for varying time $t_\mathrm{h}$, then expanded and imaged. The blue-detuned beam had a $1/e^2$ radius of 10 $\mu$m and was directed axially through the center of the condensate, as shown in \fref{exp_fig3}. During the modulation time the optical potential $U(t)$ generated by the beam followed
\begin{equation}
U(t) = U_0 \sin^2(\omega_\mathrm{mod} t/2)
\end{equation}
for $0 < t < t_\mathrm{mod}$, with maximum repulsive potential energy  $U_0$, and frequency $\omega_\mathrm{mod}$.  For all other times, the beam was turned completely off.  We fixed the modulation time at integer multiples of the modulation period, $\tau_\mathrm{mod} = 2\pi / \omega_\mathrm{mod}$, such that the intensity of the blue-detuned beam always started and ended at zero, and we did not have to be concerned with ramping off the blue-detuned beam for imaging. As in the laser pulse technique, the blue-detuned beam acted as a perturbation to the confining potential.  \Fref{660mod_fig1} shows a sequence of images for varying values of $t_\mathrm{h}$ after modulating for $t_\mathrm{mod} = 187.5$ ms, with $U_0 \sim 0.3\mu_0$ (where $\mu_0 \sim 8\hbar\omega_\mathrm{z}$), and frequency $\omega_\mathrm{mod} = 2\pi \times 16$ Hz.  This was approximately twice the radial trap frequency.  An \textit{in situ} image of the BEC in the harmonic trap with the blue-detuned beam aligned in the center is shown in the leftmost image of \fref{660mod_fig2}.

While this method generated a large number of vortex cores, we also observed large-scale breathing oscillations in the radial dimension of the condensate. The radius of the expanded cloud oscillated between $\sim$ 40 $\mu$m and $\sim$ 180 $\mu$m with a period of $\sim$ 70 ms and eventually damped out with an exponential decay time constant $\tau_\mathrm{damp} \sim$ 500 ms. As shown in \fref{660mod_fig2}, a disordered distribution of vortex cores remained in the condensate even after the bulk oscillations subsided, with $\sim 8$ cores remaining for $t_\mathrm{h}$ = 2 s. It is not entirely surprising that we induced bulk fluid oscillations, given that we were forcing the BEC at a frequency that was twice the radial trapping frequency.  However, further study is needed for positively identifying the vortex generation mechanism in this experiment.  

\begin{figure}
\begin{center}
\includegraphics[width=.7\linewidth]{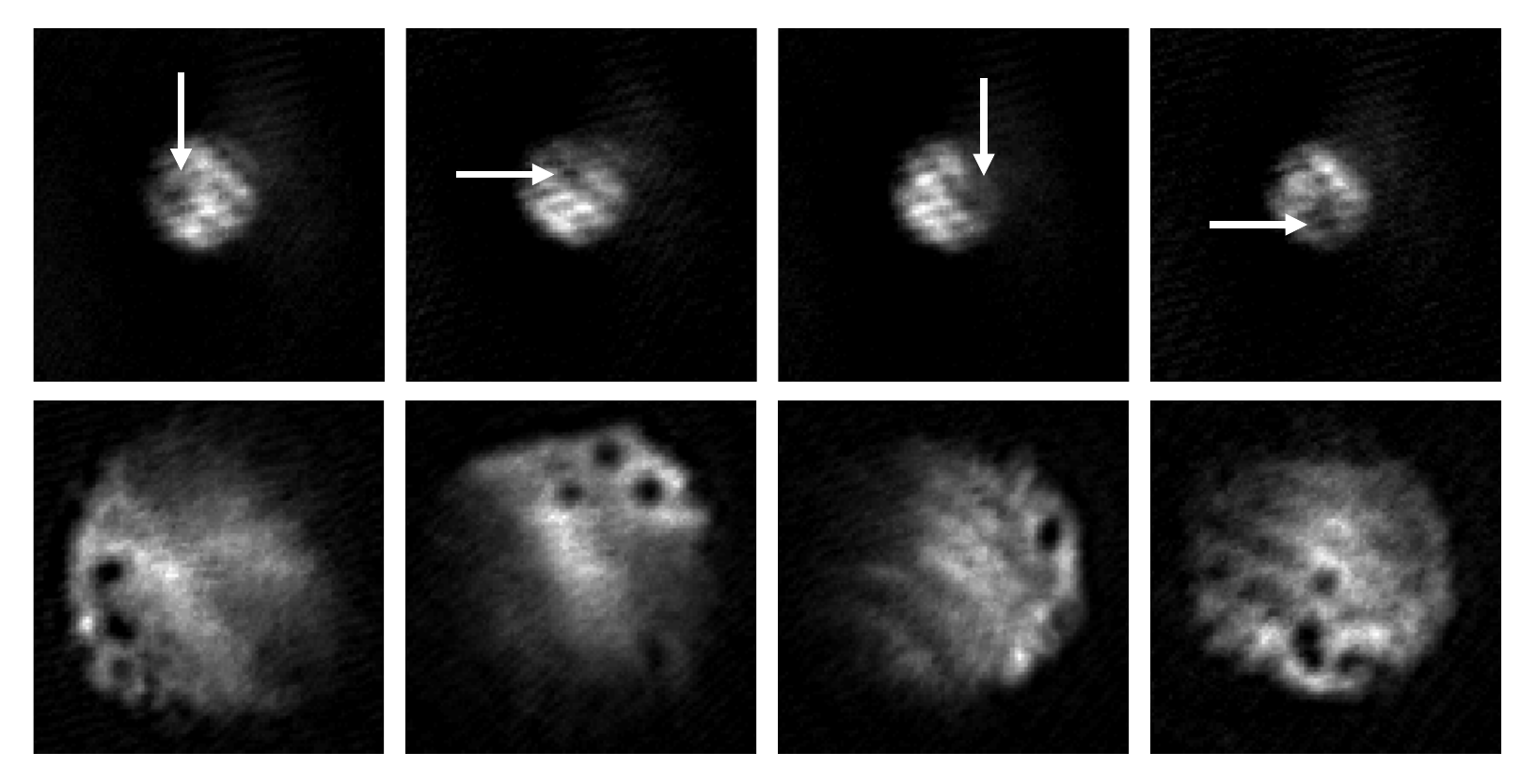}
\caption{Localization of vortex cores at the location of the blue-detuned beam. All images are 250-$\mu$m-square absorption images. The top row of images were taken with the BEC in the harmonic trap and the maximum height of the blue-detuned potential at $U_0 \sim 0.3\mu_0$.   The locations of the laser beam correspond to the positions designated by the arrow.  The bottom row of images were acquired after $t_\mathrm{h}$ = 40 ms followed by the expansion procedure. Note the correlation between beam position and the position of the vortex cores in each vertical pair of images.} \label{660mod_fig3}
\end{center}
\end{figure}

\begin{figure}[b]
\begin{center}
\includegraphics[width=.7\linewidth]{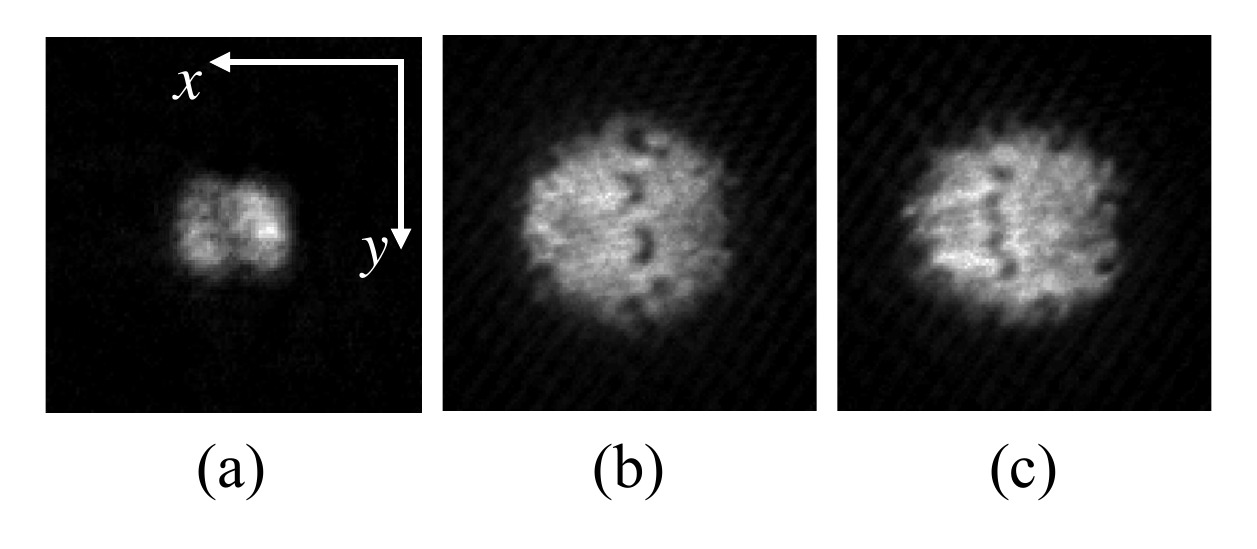}
\caption{Localization of vortex cores at the location of the blue-detuned beam. All images are 250-$\mu$m-square absorption images. (a) taken with the BEC in the harmonic trap and the blue-detuned potential on at a strength of $U_0 \sim 0.5\mu_0$. In (b) and (c), expansion images were taken after $t_\mathrm{h}$ = 40 ms. For these images  $U_0 \sim 0.3\mu_0$. Note that the vortex cores are localized along the long axis of the blue-detuned beam and that they appear to be nucleating in pairs.} \label{660mod_fig4}
\end{center}
\end{figure}

In the sequence of images shown in \fref{660mod_fig1}, it is unclear where vortex nucleation occurs.  To further explore the nucleation mechanism, we aligned the blue-detuned beam near the edge of the BEC, and modulated for $t_\mathrm{mod} = 62.5$ ms, equivalent to one sinusoidal pulse of the laser light, with $U_0 \sim 0.3\mu_0$, and frequency $\omega_\mathrm{mod} = 2\pi \times 16$ Hz. After modulation, we held the BEC in the purely harmonic trap for $t_\mathrm{h} = 40$ ms, then expanded and imaged the BEC. As shown in \fref{660mod_fig3}, the vortex cores that resulted from the modulation appear to form near the location of the focused laser beam.

\begin{figure}[t]
\begin{center}
\includegraphics[width=.8\linewidth]{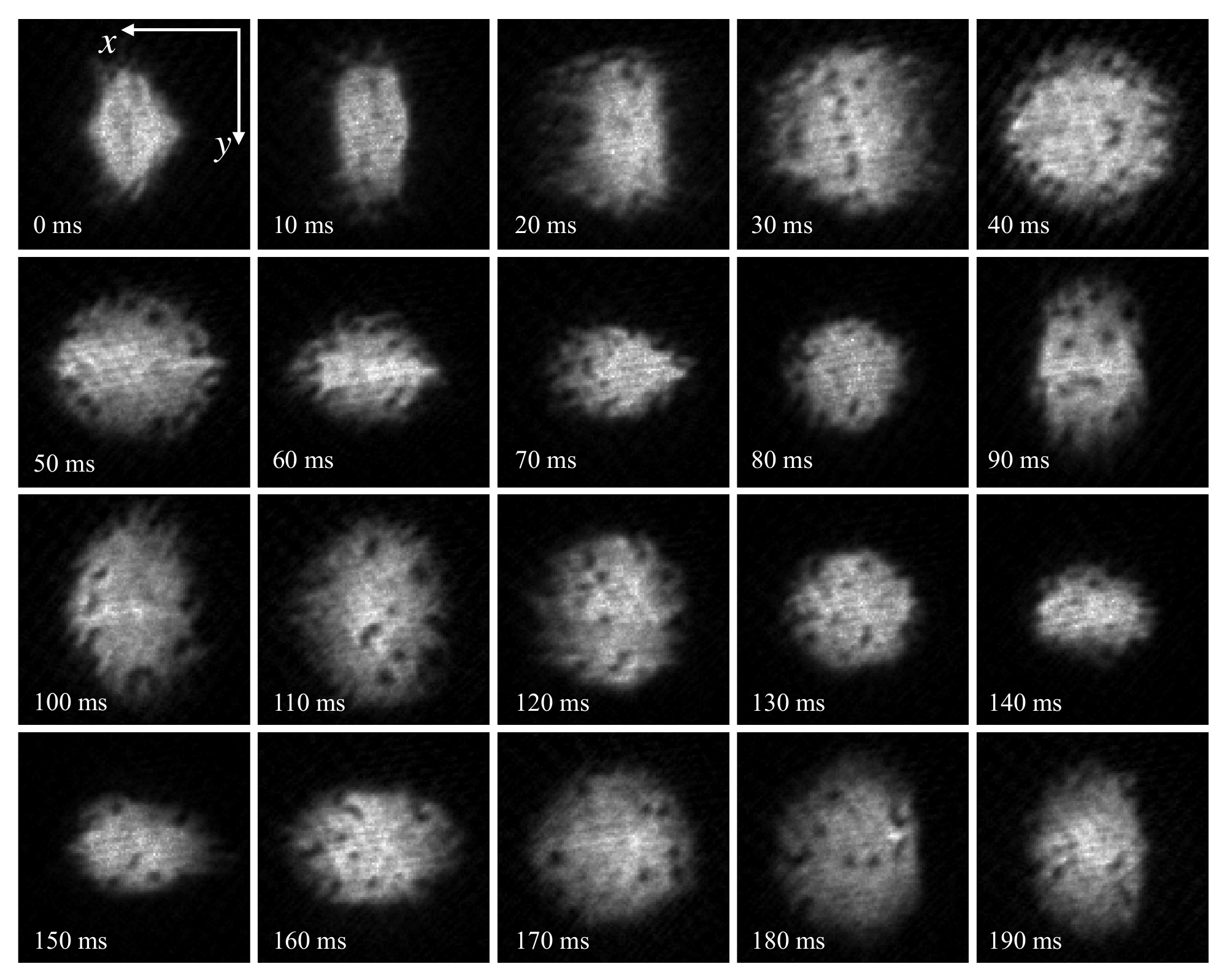}
\caption{250-$\mu$m-square absorption images acquired after the hold times $t_\mathrm{h}$ indicated. The condensate undergoes bulk oscillations that are out of phase in $x$ and $y$. Vortex cores appear to nucleate at the location of the beam and then move throughout the BEC.} \label{660mod_fig5}
\end{center}
\end{figure}

Lastly, we replaced the axially symmetric $(w_{0\mathrm{x}} = w_{0\mathrm{y}})$ focused blue-detuned Gaussian beam with a blue-detuned light sheet, focused along the $x$-axis with a $1/e^2$ radius $w_{0\mathrm{x}} = 10 ~\mu$m, spatially extended along the $y$ direction, and directed axially (along $z$) through the center of the BEC. The width of the beam along the $y$-axis was much larger than the diameter of the condensate so that beam extended beyond the edge of the condensate in the $y$-direction.  \Fref{660mod_fig4}(a) is a vertical absorption image of the unexpanded highly oblate BEC with the elongated blue-detuned beam partially penetrating the condensate. We modulated at $\omega_\mathrm{mod} = 2\pi \times 16$ Hz for $t_\mathrm{mod}$ = 62.5 ms and $U_0 \sim 0.3\mu_0$, held for $t_\mathrm{h}$, then expanded and imaged the BEC.  Figures~\ref{660mod_fig4}(b) and \ref{660mod_fig4}(c) were taken after $t_\mathrm{h}$ = 40 ms. Here the vortices appear to be nucleated along the long axis of the elongated beam. The vortex cores shown in these images are not completely resolved but we suspect that the vortices are being nucleated as dipoles in a similar manner to the breakdown of a soliton due to the snake instability in a BEC\cite{Feder2000,Anderson2001,Dut2001.Sci293.663}. Alternatively, these features may be acoustic precursors to vortex dipole formation \cite{Nazarenko07a}.

As with the axially symmetric blue-detuned potential, we observed bulk excitations in the condensate. \Fref{660mod_fig5} shows a sequence of images taken with $\omega_\mathrm{mod} = 2\pi \times 16$ Hz, $t_\mathrm{mod}$ = 62.5 ms, and $U_0 \sim 0.5\mu_0$, for varying hold times $t_\mathrm{h}$. The oscillations in $x$ and $y$ are now out of phase by $\sim90^{\circ}$. Vortex cores appear to be generated at the location of the beam and move to the outer boundary. In particular the images corresponding to $t_\mathrm{h} < 40$ ms show vortices aligned with the long axis of the beam, but by $t_\mathrm{h}$ = 50 ms the vortices are located along the outer boundary of the condensate with no cores in the center. Later images show more disordered distributions of cores. Again, the mechanism for nucleation is not clear.

In all of these methods, intensity  modulations of a blue-detuned laser beam were observed to be effective for nucleating vortices in a BEC.  Although the mechanisms for vortex nucleation remain unclear, we observe correlations between the position of the laser beam and the site of vortex generation within the BEC.  With careful parameter selection, it may be possible to use this method for controlled vortex generation rates.  However, one must also be careful not to significantly excite shape oscillations, as these would make studies of 2DQT difficult.  Perhaps with further adjustment of the modulation rates and times, shape oscillations could be minimized, and methods of this sort could be utilized for 2DQT studies.


\section{Stirring with a Blue-Detuned Laser Beam}\label{KW_sec6_3}

Rather than using intensity modulation of a blue-detuned laser beam,  the experiment described in Neely \textit{et al} \cite{Neely2012} used small-scale stirring of the BEC with a blue-detuned Gaussian beam to generate disordered vortex distributions in a highly oblate annular trapping potential.  It was also observed that vortices coalesced into large-scale flow in an annular trap, but here we review only the vortex distributions observed via stirring.   

\begin{figure}
\begin{center}
\includegraphics[width=.65\linewidth]{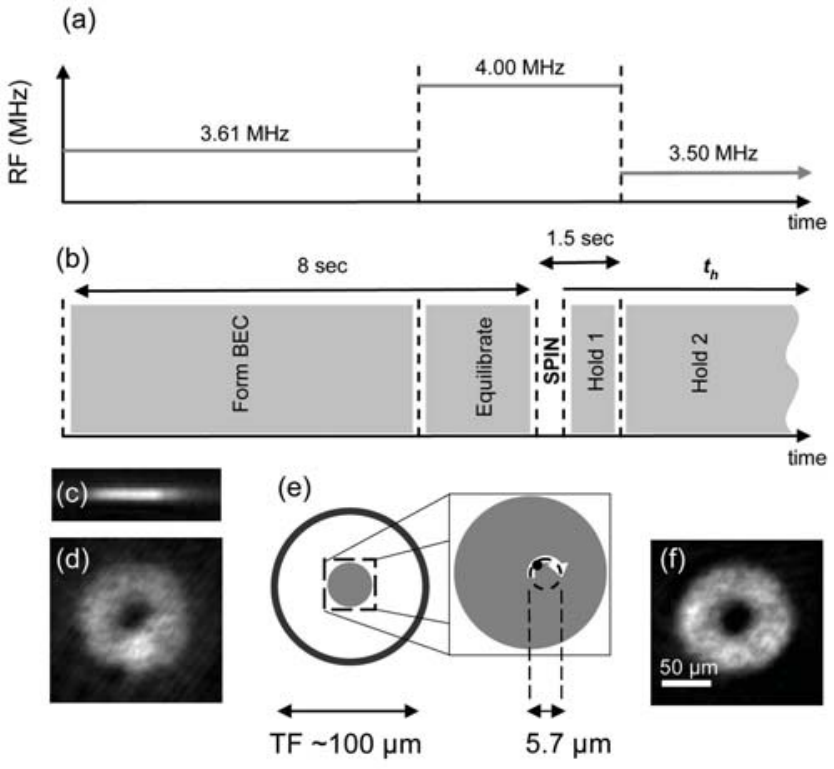}
\caption{Timing sequence used to study 2DQT. (a) and (b), experimental parameters vs.~ time.  In (a), the radio frequency (RF) field is adjusted so that the RF value is jumped further away from cutting into the BEC while the BEC is stirred with the laser beam.  After the stirring period and a 1.17-ms hold, the RF cuts into the cloud to lower the temperature of the system.  (b) illustration of the sequence of events during BEC stirring.  (c) and (d), experimental
\emph{in situ} column-density images of the BEC immediately prior
to the stir, as viewed (c) in the plane of 2D trapping and (d) along the
$z$ axis. Lighter grayscale shades indicate larger column densities. (e) illustration
of stirring, the arrow shows the trajectory of the harmonic
trap center relative to the larger fluid-free region created by the laser
barrier. (f) \emph{in situ} image of the BEC 10 s after stirring; vortices are
not observable, necessitating an expansion stage to resolve them.
} \label{current_fig1}
\end{center}
\end{figure}

This experiment was performed with the BEC in an annular trap created with a focused blue-detuned Gaussian beam, with a $1/e^2$ radius of 23 $\mu$m,  directed axially through the center of the BEC. As shown in \fref{current_fig1} the blue-detuned beam penetrated the BEC with a barrier height of $U_0 \sim 1.5\mu_0$ where $\mu_0 \sim 8\hbar\omega_\mathrm{z}$.  Prior to stirring, the BEC was held at a temperature $T \sim 0.9 T_\mathrm{C}$, where $T_\mathrm{C} \sim 116$ nK was the critical temperature for the BEC phase transition.

At $t = 0$ magnetic bias coils were used to move the center of the harmonic trap in a 5.7-$\mu$m-diameter, off-center circle about the stationary, blue-detuned barrier. At $t = 0.333$ s, at the end of the stirring motion, the center of the harmonic trap again coincided with the blue-detuned beam, and the BEC was held in this annular trap for varying hold times up to $t_\mathrm{h} \sim 50$ s. At $t_\mathrm{h}$ = 1.17 s the temperature of the BEC was reduced to $\sim 0.6 T_\mathrm{c}$ to decrease thermal damping rates and vortex-antivortex recombination. At the end of the hold time the blue-detuned beam was ramped off over 250 ms and the BEC was expanded and imaged. 

\begin{figure}
\begin{center}
\includegraphics[width=.75\linewidth]{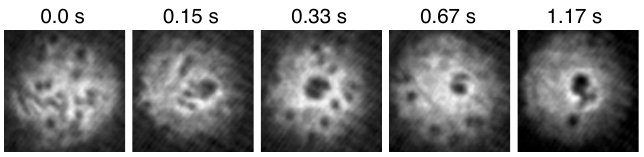}
\caption{200-$\mu$m-square experimental column-density absorption images
acquired at the hold times $t_\mathrm{h}$ indicated above the images. Each
BEC undergoes ballistic expansion immediately after the central barrier
ramp-down in order to resolve the vortex cores. Each image is
acquired from a separate experimental run. } \label{current_fig2}
\end{center}
\end{figure}

Experimental images acquired at varying hold times are shown in \fref{current_fig2}. Immediately after stirring, a large disordered distribution of vortex cores was observed.  By $t_\mathrm{h}$ = 0.33 s, vortices had begun to coalesce on the central potential barrier, as indicated by the large-vorticity hole in the center of the images.  Large-scale superflow developed over increasing hold times, as indicated by the growing large area of vorticity at the center of the BEC, and was observed to persist up to 50 s.  The initial vortex distribution generated by the stirring varied significantly from shot to shot but consistently evolved into a large-scale flow.

The damped GPE was also used to numerically simulate the BEC stirring and subsequent dynamics; see Ref.~\cite{Neely2012} for details.   In addition to the evolution of the large-scale flow observed in both experiment and numerics, analysis of the numerical simulations yielded two other results indicative of 2D turbulence. First, coherent vortex structures consisting of two cores of the same circulation intermittently formed and lasted for long timescales, tantalizing evidence for Onasger-type vortex aggregates.  Second, analysis of the incompressible kinetic energy spectrum gave $E(k) \propto k^{-5/3}$ for $k < k_i$ and $E(k) \propto k^{-3}$ for $k > k_i$ where $k_i$ is the wavenumber associated with injection of kinetic energy into the BEC.  In this case the injection mechanism was energy transfer between the sound field associated with the compressible component of the fluid excited by stirring, and the vorticity field associated with the incompressible component.   Further relationships between stirring mechanisms and observed vortex distributions and dynamics are found in Refs.~\cite{Neely2012} and \cite{Bradley2012}.  With the development of new experimental techniques, we hope that such information will soon also be experimentally discernible.  As it stands, such small-scale stirring with a blue-detuned laser beam appears to be a promising mechanism for further studies of 2DQT, perhaps even with multiple beam sites or with longer stirring times.


\section{Conclusions}\label{KW_sec7}

We have found that it is relatively straightforward to nucleate large, disordered distributions of vortex cores in highly oblate BECs, that is, to generate 2DQT in a BEC.  We interpret the disordered nature of these distributions as analogous to the vortex tangle characteristic of 3D quantum turbulence.  Many of the vortex distributions that we have observed are long-lived, making them potentially useful for studies of 2DQT. However, methods such as modulating a blue-detuned beam in the center of the BEC excite bulk modes of the fluid where the entire condensate undergoes shape oscillations.  In many cases, the vortex nucleation mechanisms are not completely clear.  Possible mechanisms for vortex generation in the laser modulation experiments include counterflow between the thermal and superfluid components \cite{QVDSF,Vin1957.PRSLA242.493,Vin1958.PRSLA243.400}, vacuum bubble cavitation as described by Berloff and Barenghi \cite{Berloff2004}, or perhaps other nonlinear fluid dynamics processes at the boundaries of the quantum fluid.  While further numerical and experimental investigations are necessary to determine the origins of the vortices, we envision many of these highly disordered vortex distributions as starting points for further studies of 2DQT.

Although we have described a number of vortex generation mechanisms, a primary challenge for experimentalists is to characterize 2DQT in a BEC, and the characteristics observed may well depend on the vortex generation mechanism used.  Experimental observation of an inverse energy cascade, vortex aggregation, and kinetic energy spectra are of primary interest.  Such characteristics might or might not appear via the methods discussed in this article; this remains to be determined.  Vortex dynamics and circulation measurements will likely prove to be among the most significant developments remaining to be developed for 2DQT studies in a BEC.  Understanding the temperature dependence, and more generally the role of dissipation, are also of high importance.  Finally, reaching a regime of quasi-steady-state 2DQT is a highly desired goal, one that may well be reached utilizing methods described in this article, presumably with a suitable balance of dissipation.  

Although many of the methods we describe in this article represent only the beginnings of investigations into new methods for the study of 2DQT, we hope that the descriptions of our observations so far inspire the development of new experimental and numerical studies of vortex generation mechanisms.  Further experimental and theoretical work in the field of 2DQT promises to provide an exciting new research direction for compressible quantum fluids.  \\
\\
We gratefully acknowledge the support of the US National Science Foundation grant PHY-0855467.  KEW gratefully acknowledges support from the Department of Energy Office of Science Graduate Fellowship Program (DOE SCGF), made possible in part by  the American Recovery and Reinvestment Act of 2009, administered by ORISE-ORAU under contract no.~DE-AC05-06OR23100.   We also thank Ashton Bradley and Ewan Wright for helpful discussions.

\end{document}